\documentclass[ALICE,manyauthors]{cernphprep}
\usepackage[comma,square,numbers,sort&compress]{natbib}
\usepackage{hyperref}
\usepackage{lineno}
\usepackage{xspace}
\usepackage{color}
\usepackage{amsmath}
\usepackage{relsize}
\usepackage[T1]{fontenc}
\usepackage{orcidlink}

\begin{document}

\newcommand{\jpsi}{\ensuremath{\rm J/\psi}\xspace}
\newcommand{\ccBar}{\ensuremath{c\bar{c}}\xspace}
\newcommand{\bbBar}{\ensuremath{b\bar{b}}\xspace}
\newcommand{\mee}{\ensuremath{m_{\rm{ee}} }\xspace}

%

\newcommand{\pp}           {pp\xspace}
\newcommand{\ppbar}        {\mbox{$\mathrm {p\overline{p}}$}\xspace}
\newcommand{\XeXe}         {\mbox{Xe--Xe}\xspace}
\newcommand{\PbPb}         {\mbox{Pb--Pb}\xspace}
\newcommand{\pA}           {\mbox{pA}\xspace}
\newcommand{\pPb}          {\mbox{p--Pb}\xspace}
\newcommand{\AuAu}         {\mbox{Au--Au}\xspace}
\newcommand{\UU}         {\mbox{U--U}\xspace}
\newcommand{\dAu}          {\mbox{d--Au}\xspace}

\newcommand{\s}            {\ensuremath{\sqrt{s}}\xspace}
\newcommand{\snn}          {\ensuremath{\sqrt{s_{\mathrm{NN}}}}\xspace}
\newcommand{\pt}           {\ensuremath{p_{\rm T}}\xspace}
\newcommand{\meanpt}       {$\langle p_{\mathrm{T}}\rangle$\xspace}
\newcommand{\ycms}         {\ensuremath{y_{\rm CMS}}\xspace}
\newcommand{\ylab}         {\ensuremath{y_{\rm lab}}\xspace}
\newcommand{\etarange}[1]  {\mbox{$\left | \eta \right |~<~#1$}}
\newcommand{\yrange}[1]    {\mbox{$\left | y \right |~<~#1$}}
\newcommand{\dndy}         {\ensuremath{\mathrm{d}N_\mathrm{ch}/\mathrm{d}y}\xspace}
\newcommand{\dndeta}       {\ensuremath{\mathrm{d}N_\mathrm{ch}/\mathrm{d}\eta}\xspace}
\newcommand{\avdndeta}     {\ensuremath{\langle\dndeta\rangle}\xspace}
\newcommand{\dNdy}         {\ensuremath{\mathrm{d}N_\mathrm{ch}/\mathrm{d}y}\xspace}
\newcommand{\Npart}        {\ensuremath{N_\mathrm{part}}\xspace}
\newcommand{\Ncoll}        {\ensuremath{N_\mathrm{coll}}\xspace}
\newcommand{\dEdx}         {\ensuremath{\textrm{d}E/\textrm{d}x}\xspace}
\newcommand{\RpPb}         {\ensuremath{R_{\rm pPb}}\xspace}

\newcommand{\nineH}        {$\sqrt{s}~=~0.9$~Te\kern-.1emV\xspace}
\newcommand{\seven}        {$\sqrt{s}~=~7$~Te\kern-.1emV\xspace}
\newcommand{\twoH}         {$\sqrt{s}~=~0.2$~Te\kern-.1emV\xspace}
\newcommand{\twosevensix}  {$\sqrt{s}~=~2.76$~Te\kern-.1emV\xspace}
\newcommand{\five}         {$\sqrt{s}~=~5.02$~Te\kern-.1emV\xspace}
\newcommand{\twosevensixnn}{$\sqrt{s_{\mathrm{NN}}}~=~2.76$~Te\kern-.1emV\xspace}
\newcommand{\fivenn}       {$\sqrt{s_{\mathrm{NN}}}~=~5.02$~Te\kern-.1emV\xspace}
\newcommand{\LT}           {L{\'e}vy-Tsallis\xspace}
\newcommand{\GeVc}         {Ge\kern-.1emV/$c$\xspace}
\newcommand{\MeVc}         {Me\kern-.1emV/$c$\xspace}
\newcommand{\TeV}          {Te\kern-.1emV\xspace}
\newcommand{\GeV}          {Ge\kern-.1emV\xspace}
\newcommand{\MeV}          {Me\kern-.1emV\xspace}
\newcommand{\GeVmass}      {Ge\kern-.2emV/$c^2$\xspace}
\newcommand{\MeVmass}      {Me\kern-.2emV/$c^2$\xspace}
\newcommand{\lumi}         {\ensuremath{\mathcal{L}}\xspace}

\newcommand{\ITS}          {\rm{ITS}\xspace}
\newcommand{\TOF}          {\rm{TOF}\xspace}
\newcommand{\ZDC}          {\rm{ZDC}\xspace}
\newcommand{\ZDCs}         {\rm{ZDCs}\xspace}
\newcommand{\ZNA}          {\rm{ZNA}\xspace}
\newcommand{\ZNC}          {\rm{ZNC}\xspace}
\newcommand{\SPD}          {\rm{SPD}\xspace}
\newcommand{\SDD}          {\rm{SDD}\xspace}
\newcommand{\SSD}          {\rm{SSD}\xspace}
\newcommand{\TPC}          {\rm{TPC}\xspace}
\newcommand{\TRD}          {\rm{TRD}\xspace}
\newcommand{\VZERO}        {\rm{V0}\xspace}
\newcommand{\VZEROA}       {\rm{V0A}\xspace}
\newcommand{\VZEROC}       {\rm{V0C}\xspace}
\newcommand{\Vdecay} 	   {\ensuremath{V^{0}}\xspace}

\newcommand{\ee}           {\ensuremath{e^{+}e^{-}}} 
\newcommand{\pip}          {\ensuremath{\pi^{+}}\xspace}
\newcommand{\pim}          {\ensuremath{\pi^{-}}\xspace}
\newcommand{\kap}          {\ensuremath{\rm{K}^{+}}\xspace}
\newcommand{\kam}          {\ensuremath{\rm{K}^{-}}\xspace}
\newcommand{\pbar}         {\ensuremath{\rm\overline{p}}\xspace}
\newcommand{\kzero}        {\ensuremath{{\rm K}^{0}_{\rm{S}}}\xspace}
\newcommand{\lmb}          {\ensuremath{\Lambda}\xspace}
\newcommand{\almb}         {\ensuremath{\overline{\Lambda}}\xspace}
\newcommand{\Om}           {\ensuremath{\Omega^-}\xspace}
\newcommand{\Mo}           {\ensuremath{\overline{\Omega}^+}\xspace}
\newcommand{\X}            {\ensuremath{\Xi^-}\xspace}
\newcommand{\Ix}           {\ensuremath{\overline{\Xi}^+}\xspace}
\newcommand{\Xis}          {\ensuremath{\Xi^{\pm}}\xspace}
\newcommand{\Oms}          {\ensuremath{\Omega^{\pm}}\xspace}
\newcommand{\degree}       {\ensuremath{^{\rm o}}\xspace}

\begin{titlepage}
\PHyear{2024}       
\PHnumber{232}      
\PHdate{09 September}  

\title{Coherent \jpsi photoproduction at midrapidity in \PbPb collisions at $\pmb{\sqrt{s_{\rm NN}} = 5.02}$ TeV}
\ShortTitle{Coherent \jpsi photoproduction in \PbPb collisions at 5.02 TeV}   

\Collaboration{ALICE Collaboration\thanks{See Appendix~\ref{app:collab} for the list of collaboration members}}
\ShortAuthor{ALICE Collaboration} 

\begin{abstract}

The coherent \jpsi photoproduction cross section is 
measured for the first time at midrapidity in peripheral to semicentral
\PbPb collisions at \fivenn.  The centrality differential cross
section \mbox{${\rm d} \sigma/ {\rm d}y$} is reported for the centrality range
40--90\%, together with the doubly-differential cross section
\mbox{${\rm d}^2 \sigma /{\rm d}y {\rm d} p_{\rm T}$}, extracted in two peripheral centrality classes.
The \jpsi mesons are reconstructed in the dielectron channel, in the
rapidity interval \mbox{$|y| <$ 0.9} using the ALICE central barrel
detectors. 
The \jpsi cross section at midrapidity is statistically compatible to the earlier ALICE measurement at forward rapidity and at the same centre-of-mass energy, and shows only a mild centrality dependence over the covered range.
Several sets of theoretical calculations 
taking into account the hadronic overlap in the collisions but ignoring
possible final-state effects from a hot expanding medium are found to
give a fairly good description of the current measurements within uncertainties. 

\end{abstract}
\end{titlepage}

\setcounter{page}{2} 


\section{Introduction} 
\label{sec:intro}

In ultrarelativistic collisions between heavy ions, the Lorentz-contracted nuclei generate strong electromagnetic fields, and either of them can be treated as the source of a quasi-real photon, which interacts with the other nucleus treated as the target. 
The photoproduction cross section can be factorised as the product of
the photon flux emanating from one of the colliding nuclei, which is
proportional to the square of the nuclear electric charge, $Z^2$, and the photonuclear
cross section.
This process
can be viewed as a fluctuation of the photon into a quark-antiquark pair (colour dipole), which then interacts with the gluon field of the target nucleus, at leading order through the exchange of two gluons in a colour singlet state, producing a real vector meson~\cite{Klein:2019qfb}. In this approximation, the cross section is proportional to the square of the target gluon distribution at relevant values of ($x$, $Q^2$)~\cite{Baltz:2007kq}.
The photoproduction can be either coherent, with the photon interacting with the target nucleus as a whole, generating a vector meson with very low transverse momentum (\mbox{$\langle p_{\rm T} \rangle \sim$ 60 MeV/$c$}), or incoherent, with the photon interacting with a single nucleon, producing a meson of higher transverse momentum (\mbox{$\langle p_{\rm T} \rangle \sim$ 500 MeV/$c$})~\cite{Bertulani:2005ru}. Photonuclear vector meson production is thus a powerful tool for investigating the gluonic structure of the target.
Cross section measurements are sensitive to the gluon density in the nucleus and, due to the relatively low values of $x$ that are accessible via this process, may constitute a probe for gluon shadowing or saturation. In addition, spatial characteristics of the gluon distribution, such as transverse size, shape, and fluctuations, can be accessed through measurements of the differential cross section \mbox{$d \sigma/ dt$}, with the Mandelstam variable \mbox{$t \sim -p_{\rm T}^2$}~\cite{ALICE:2021tyx}. 

Photonuclear reactions have been most extensively studied in ultraperipheral
collisions (UPCs) of heavy ions, taking place at impact parameters larger
than the sum of their radii.  
With combinatorial and hadronic backgrounds being negligible,
coherent and incoherent cross sections can readily be extracted. 
At the LHC energies, ALICE, CMS, and LHCb have studied coherent charmonium
photoproduction in ultraperipheral \PbPb collisions at
\snn = 2.76 TeV~\cite{ALICE:2012yye, ALICE:2013wjo, CMS:2016itn} and
5.02 TeV~\cite{ALICE:2019tqa, ALICE:2021gpt, ALICE:2021tyx, LHCb:2021bfl, ALICE:2023jgu, CMS:2023snh}, 
in the rapidity ranges
\mbox{$|y| <$ 0.9} (ALICE), \mbox{1.8 $< |y| <$ 2.3 (CMS, 2.76 \TeV)}, \mbox{1.6 $< |y| <$ 2.4 (CMS, 5.02 \TeV)}, \mbox{2.5 $< y <$ 4} (ALICE),
and \mbox{2.0 $< y <$ 4.5} (LHCb). At the RHIC energies, the coherent and incoherent charmonium photoproduction cross sections in ultraperipheral Au-Au collisions at \snn=200 GeV 
have recently been reported by STAR~\cite{STAR:2023nos}. At both the LHC and RHIC, the measurements indicate a sizable suppression compared to the
expectations from the impulse approximation. These observations are qualitatively reproduced by both gluon shadowing and gluon saturation phenomenological models,
but none of them reproduce these data quantitatively over the entire covered Bjorken-$x$ region.

Coherent photoproduction of \jpsi in heavy-ion interactions with nuclear overlap, manifesting as a prominent excess (for \mbox{$p_{\rm T} <$ 0.3 GeV/$c$}) relative to the soft tail of the hadronically generated \jpsi spectrum, was first observed by ALICE in peripheral \PbPb collisions at \mbox{\snn = 2.76 TeV} and at forward rapidity \mbox{(2.5 $< y <$ 4)}, in the centrality range 30--90\%~\cite{ALICE:2015mzu}. The phenomenon was confirmed by STAR at RHIC energies, in \AuAu (20--80\%) and \UU (40--80\%) collisions at \snn = 200 and 193 GeV, respectively, also presenting the $p_{\rm T}$ and $|t|$ dependence of the coherent cross section~\cite{STAR:2019yox}. 
Recently, LHCb~\cite{LHCb:2021hoq} and ALICE~\cite{ALICE:2022zso} reported a similar excess also in \PbPb collisions at \snn = 5.02 TeV, in both cases measured at forward rapidity (2 $< y <$ 4.5 and 2.5 $< y <$ 4, respectively).
LHCb, covering a range of peripheral collisions (approximately 65--90\%), measured the differential \jpsi photoproduction yield ${\rm d}N/{\rm d}y$ as a function of the number of participants $\langle N_{\rm part} \rangle$ and $y$, and the doubly-differential yield ${\rm d}^2 N/{\rm d}y {\rm d}p_{\rm T}$ as a function of $p_{\rm T}$.  ALICE extracted the coherent \jpsi photoproduction centrality differential cross section ${\rm d} \sigma/{\rm d}y$ 
over the centrality range 0--90\%, setting an upper limit for the most central collisions (0--10\%). A mild centrality dependence is revealed, 
compatible within uncertainties
both with a flat trend and with a decrease of the cross section towards
more central collisions.

The observation of coherent photoproduction in nuclear collisions with overlap inspired new theoretical developments~\cite{GayDucati:2018who, Cepila:2017nef, Zha:2017jch, Zha:2018jin, Klusek-Gawenda:2015hja, Shi:2017qep, Jenkovszky:2022wcw}, attempting to quantitatively describe the centrality, rapidity, and in the case of Refs.~\cite{Zha:2017jch, Zha:2018jin, Jenkovszky:2022wcw}, the $p_{\rm T}$ dependence of the experimentally measured cross sections or yields. In order to account for the hadronic overlap, these models take the UPC picture as a baseline and impose geometric constraints implemented as impact parameter ranges, modifying the photon flux and, in some cases, also the photonuclear cross section.

Several questions remain unresolved regarding coherent photoproduction processes in a hadronic environment, including the roles of spectator and participant nucleons as photon sources and targets (theoretically explored in~\cite{Zha:2017jch}), the survival of the coherence condition in the presence of nuclear breakup, and the time ordering of the hadro- and photoproduction. The coherent \jpsi $p_{\rm T}$ distribution, its impact parameter dependence, and the possible influence of destructive interference between the two photon sources and of the strong interactions in the nuclear overlap zone have been investigated in Refs.~\cite{Zha:2017jch, Zha:2018jin}. The shapes of the measured $|t|$ spectrum in Ref.~\cite{STAR:2019yox} and $p_{\rm T}$ spectrum in Ref.~\cite{LHCb:2021hoq} are well reproduced by these calculations within uncertainties. However, no available experimental data allow meaningful assessment of the evolution of $p_{\rm T}$ distributions with collision centrality. 

An issue still sparsely explored by models is the potential influence of the hot and rapidly expanding partonic medium generated in the hadronic overlap zone 
on the coherently photoproduced, low-\pt charmonia. 
The latter are formed during the same narrow
time interval as the initial hadronic interactions occur, presumably
distributed over the surface of the target nuclei~\cite{Shi:2017qep},
and remain almost stationary in the transverse plane.
This subset of the observable \jpsi
population is uniquely identifiable as primordial
survivors through their characteristic \pt distribution, and is, therefore,
a particularly interesting QGP probe.
Possible final-state medium 
effects are expected to be nearly absent in the most peripheral collisions,
but may exhibit an onset with increasing nuclear overlap, manifesting
as a reduction in measured cross sections and possibly a modulation
in azimuthal distributions beyond that predicted by models considering
only geometric constraints on photoproduction. Taking into account gluon-induced dissociation of charmonia in the plasma, Ref.~\cite{Shi:2017qep}
predicts a medium-induced suppression of the photoproduced \jpsi yield
of $\sim$ 20--40\% for semicentral collisions with
$\langle N_{\rm part} \rangle$ $\sim$ 100–150.

This paper presents for the first time the \jpsi coherent photoproduction cross sections, measured at midrapidity in peripheral to semicentral (40--90\%) \PbPb collisions at \snn = 5.02 TeV.
The doubly-differential cross section
${\rm d}^2 N/{\rm d}y {\rm d}p_{\rm T}$ extracted by ALICE is reported for the first
time for \jpsi photoproduction in collisions with nuclear overlap for two centrality classes, 50--70\% and 70--90\%.
Section 2 gives an overview of the experimental apparatus
and the data sample used, while Section 3 details the analysis
and the extraction of systematic uncertainties.  Section 4 presents the results
and discusses them in the context of other existing \jpsi photoproduction
measurements and relevant model calculations.  Finally, Section 5
summarises the conclusions based on the LHC Run 2 data and points to opportunities
offered by upcoming high-luminosity runs.

\section{Experimental setup and data sample}
\label{sec:experiment}

A detailed description of the ALICE detector and its performance is provided in Refs.~\cite{Aamodt:2008zz, ALICE:2014sbx}.  
In this work, \jpsi mesons are reconstructed in the dielectron
channel at midrapidity ($|y|<0.9$), 
using the  central barrel charged particle tracking detectors ITS (Inner Tracking System) ~\cite{ALICE:2010tia} and TPC (Time Projection Chamber)~\cite{Alme:2010ke} which provide very good tracking and electron identification in the pseudorapidity range $|\eta|<0.9$.
The ITS comprises six cylindrical detection layers based on different silicon technologies, surrounding the beam axis at radii ranging from 3.9 to 43 cm.  The two innermost layers, requiring the highest granularity, are Silicon Pixel Detectors (SPD), while the two middle layers are Silicon Drift Detectors (SDD) and the two outermost ones are Silicon Strip Detectors (SSD). The ITS system is used for the determination of the event primary interaction point, precision tracking for the location of secondary vertices from weakly decaying particles, and event selection. The TPC is a cylindrical gaseous detector, filled with a mixture of argon and carbon dioxide, surrounding the ITS, with an inner radius of 0.85 m, an outer radius of 2.5 m, and a length of 5 m along the beam direction. It is the main tracking detector of the central barrel system, also performing particle identification based on the specific energy loss ${\rm d}E/{\rm d}x$ in the gas.  

In addition, a suite of detectors for global event characterisation is employed. The V0 detector~\cite{ALICE:2013axi} consists of two scintillating arrays covering the pseudorapidity ranges \mbox{2.8 $< \eta <$ 5.1} (V0A) and  \mbox{$-$3.7 $< \eta <$ $-$1.7} (V0C), both being segmented in pseudorapidity and azimuthal angle.
This subsystem is used for triggering and rejecting beam-induced background events, measuring charged particle multiplicity, and determining collision centrality and
event plane azimuthal angle. The centrality in \PbPb collisions is defined as the percentile of the hadronic cross section corresponding to the charged particle multiplicity measured in the V0 detector~\cite{ALICE:2013hur}.  A fit of a Glauber Monte Carlo model to the V0 amplitude distribution is used to relate the experimental centrality with geometric quantities like the average impact parameter ($\langle b \rangle$) and the number of participants ($\langle N_{\rm part} \rangle$). 
The Zero Degree Calorimeter~\cite{Arnaldi:1999zz} comprises
two sets of detector arrays located close to the beam axis on both
sides of the nominal interaction point, at a distance of $\pm$ 112.5 m.
Their tasks are detecting nucleons emitted at zero degree relative to 
the beam axis, aiding event characterisation both in hadronic and
electromagnetic interactions, and rejection of beam-induced background.

The analysed data set was collected by ALICE during the 2015 and 2018 LHC \PbPb runs at \mbox{\snn = 5.02}~TeV. At midrapidity, data was acquired using the minimum bias (MB) trigger, defined by the coincidence of signals in both the V0A and V0C arrays.  In the 2018 period, a central and a semicentral trigger were used in addition. These were defined using the MB trigger in combination with thresholds on the total online signal
amplitude in the V0 detector, corresponding roughly to collision centralities of 0--10\% and 30--50\%, respectively.  In this analysis, only events from the centrality range 40--90\% were considered.
Beam-induced background was rejected using timing information from the V0 and the Zero Degree Calorimeter detectors. 
All events were required to have a reconstructed primary vertex with a longitudinal position within $\pm$10 cm from the nominal interaction point. Events with pileup occurring during the TPC readout time were rejected in the offline analysis based on the correlation between the number of TPC and ITS (SDD+SSD) clusters.
After all selections, the number of analysed events is approximately 37 million, 35 million, and 35 million for the 40--50\%, 50--70\%, and 70--90\% centrality ranges, respectively. This corresponds to an integrated luminosity of $\sim$ 49.6$~\mu \rm{b}^{-1}$ for the 40--50\% centrality interval and $\sim$ 24$~\mu \rm{b}^{-1}$ for both the 50--70\% and 70--90\% centrality 
intervals~\cite{ALICE-PUBLIC-2021-001}.

\section{Data analysis}
\label{sec:analysis}

The \jpsi mesons are reconstructed using the $e^+ e^-$ decay channel. The selected electron candidates are good quality tracks reconstructed through both the ITS and TPC, with a minimum momentum ($p$) of 1 GeV/$c$ and a pseudorapidity $|\eta| <$ 0.9. To ensure excellent tracking quality, each track is required to have
a minimum number of 70 associated space points in the TPC and a maximum
calculated $\chi^2/N_{\rm dof}$ value of 2 for the fit of the track to
the clusters. Secondary particles are suppressed by requiring the maximum distance-of-closest-approach (DCA) of the track to the interaction vertex to be 1 cm in the transverse and 3 cm in the longitudinal direction.  In addition, daughters of long-lived weakly decaying particles are removed using topological selections.
In order to improve the tracking resolution and to reduce the
number of secondary electrons from photon conversions in the detector material, at least one hit in either of the two SPD layers is required. Electrons and positrons are
identified via their specific energy loss in the TPC gas by selecting a band of width $\pm 3 \sigma_e$ around the electron
expectation value estimated from a parameterisation of
the measured average ${\rm d}E/{\rm d}x$ as a function of
momentum~\cite{Blum:2008nqe}, with $\sigma_e$ being the resolution of this measurement. To further reduce the contamination from hadrons, tracks with a ${\rm d}E/{\rm d}x$ compatible with the pion or proton hypothesis within $\pm 3.5 \sigma_{\pi / p}$,
are rejected. Finally, electrons from photon conversions are further suppressed by using a prefiltering method~\cite{ALICE:2019pid} where candidate tracks forming a pair with invariant mass $m_{ee} <$ 50 MeV/$c^2$ when combined
with a set of candidates selected using looser cuts, are excluded from the analysis.  

In order to extract the yields of coherently photoproduced \jpsi,
a two-dimensional distribution $N(m_{ee}, p_{\rm T})$ is constructed from all combinations of opposite-sign electron tracks from the same event and passing all selections described above. An unbinned 2-dimensional log-likelihood fitting procedure is applied to the measured distribution using a model which includes contributions from
photoproduction ($F_{{\rm phot,i}}$), hadroproduction ($F_{\rm hadr}$), and background ($F_{\rm bkg}$):
\begin{equation}
  F(m_{ee}, p_{\rm T}) = f_{\rm bkg} \cdot F_{\rm bkg}(m_{ee}, p_{\rm T}) +
  f_{\rm hadr} \cdot F_{\rm hadr}(m_{ee}, p_{\rm T})
  + f_{\rm phot} \cdot \sum_{i} w_{{\rm phot},i} F_{{\rm phot},i} (m_{ee}, p_{\rm T})~,
  \label{eq:FitEq}
\end{equation}
where $f_{\rm bkg}$, $f_{\rm hadr}$, and $f_{\rm phot}$ are the normalisation parameters of the respective components, and $w_{{\rm phot,}i}$ are the fractional contributions from the different 
photoproduction processes, all described in more detail below.  

The background component, $F_{\rm bkg}$, is constructed as the sum of the combinatorial background and a much smaller background contribution from correlated semileptonic decays of heavy-quark pairs. The combinatorial background, $F_{\rm bkg}^{\rm comb}$, is constructed using the mixed-event technique, pairing opposite-sign electrons from events with similar global characteristics (centrality, vertex position, and event plane orientation). It is normalised using the like-sign pairs according to the expression
\begin{equation}
    F_{\rm bkg}^{\rm comb} (\mee,\pt) = N^{\rm ME-OS} (\mee,\pt) \cdot \frac{\sum_{\mee^{\rm i},\pt^{\rm j}}N^{\rm SE-LS}(\mee^{\rm i}, \pt^{\rm j})}{\sum_{\mee^{\rm i}, \pt^{\rm j}}N^{\rm ME-LS}(\mee^{\rm i}, \pt^{\rm j})},
    \label{eq:me}
\end{equation}
where $N^{\rm ME-OS}$ is the distribution of opposite-sign mixed-event pairs, while $N^{\rm SE-LS}$ and $N^{\rm ME-LS}$ are the like-sign pair distributions from the same and mixed events, respectively, summed over the \mee and \pt. More details on the mixed-event procedure can be found in Ref.~\cite{ALICE:2021dtt}.
The two-dimensional correlated background component, $F^{\rm corr}_{\rm bkg}$, is constructed using a parameterisation of the residual background obtained from the $(\mee, \pt)$ dielectron distribution by subtracting the combinatorial background. The parameterisation is a 2D function which is factorisable into a first-order polynomial over the invariant mass dimension and a piecewise polynomial function over the \pt dimension. The correlated background component has a relatively small amplitude
compared to the combinatorial background and to the \jpsi signal as well as to
statistical fluctuations.  The unbinned likelihood fit is therefore performed both with and without this
component, yielding a small difference taken as a systematic uncertainty, as described later in this
section. Due to the fact that $F^{\rm comb}_{\rm bkg}$ is already normalised via Eq.~\ref{eq:me}, $F^{\rm corr}_{\rm bkg}$ and the total sum $F_{\rm bkg}$
will also be normalised by construction, and the corresponding parameter $f_{\rm bkg}$ is therefore fixed in the fit of Eq.~\ref{eq:FitEq}.

The two-dimensional hadroproduction component, $F_{\rm hadr}$, is
estimated via a Monte-Carlo simulation, described in more detail 
in Refs.~\cite{ALICE:2019nrq,ALICE:2023gco}. This MC used a realistic kinematic distribution of
inclusive unpolarised \jpsi based on existing measurements, including 
a fit to inclusive yields reported in Ref.~\cite{ALICE:2019nrq}. The \jpsi were embedded 
in an underlying environment of \PbPb collisions generated using 
HIJING 1.0~\cite{Wang:1991hta}.  
The \pt shape of the embedded \jpsi signal was further
tuned to match the \pt differential \jpsi yields recently measured
by ALICE in the relevant centrality classes~\cite{ALICE:2023gco}.
The \jpsi decay is forced into the dielectron channel using the PHOTOS package ~\cite{Golonka:2005pn}. The simulated particles were transported
through a model of the ALICE detector using GEANT3~\cite{Brun:1119728}, and then
reconstructed with the same algorithms as used for real data.

The photoproduction component, $\Sigma_i w_{{\rm phot},i} F_{{\rm phot},i}$, is a sum of contributions from several processes, $i$, namely: coherent and incoherent \jpsi photoproduction, feed-down from coherently and incoherently photoproduced $\psi(2S)$, incoherent \jpsi photoproduction with nucleon dissociation, and continuum \mbox{$\gamma \gamma \rightarrow \ee$}.
These processes were simulated using the STARLight generator~\cite{Klein:2016yzr} with the same relative weights as those obtained in the UPC analysis reported in Ref.~\cite{ALICE:2021gpt}, and embedded into \PbPb collisions generated using the HIJING model with a similar setup as the one described for the hadronic component. Integrated in the range $\pt<0.2$~\GeVc, the ratio between the incoherent, feed-down and $\gamma\gamma$ continuum to the coherent \jpsi component is approximately 1\%, 2.5\% and 11\%, respectively.  
The charmonia generated with STARLight are transversely
polarised, as expected from $s$-channel helicity conservation and recently confirmed experimentally for coherently photoproduced \jpsi~\cite{ALICE:2023svb}.
For the incoherent \jpsi with nucleon dissociation component, a process not incorporated in STARLight, the \pt shape was constructed using the H1 parameterisation~\cite{H1:2013okq}. 
The individual two-dimensional template shapes of the various processes, $F_{{\rm phot,}i}(\mee, \pt)$, were obtained from the reconstructed electron pairs using the same reconstruction algorithm and analysis selections as for the data. The sum of all the reconstructed and weighted templates is used as a single component in the unbinned fit of Eq.~\ref{eq:FitEq} due to the fact that the size of the current data set does not allow the normalisation factors ($f_{\rm phot} \cdot 
w_{{\rm phot,}i}$) of all these templates to vary independently in the fit.

\begin{figure}[tb]
 \centering
    \includegraphics[width = 1\textwidth]{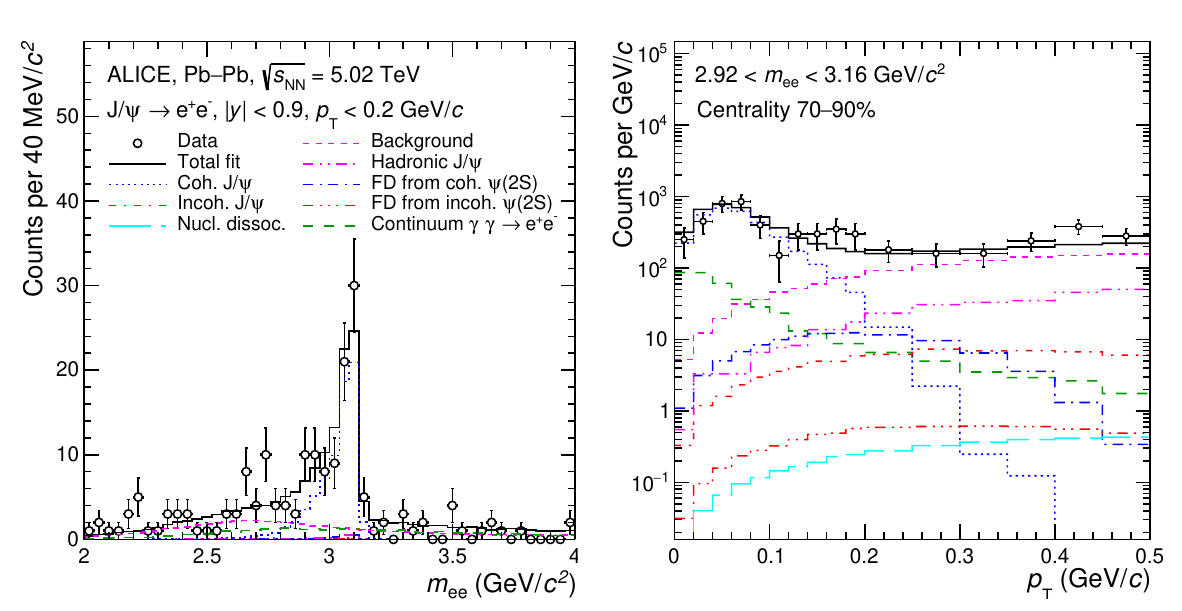}
    \caption{Invariant mass distribution for \pt $<$ 200~\MeVc (left panel) and \pt distribution for $2.92 < \mee < 3.16$~\GeVmass (right panel) in the centrality class 70--90\% for dielectrons with \mbox{$|y| < 0.9$} in \PbPb collisions at \fivenn. Measured distributions for \jpsi candidates are shown as black markers and the fit components are shown as coloured lines, as explained in the legend.}
    \label{fig:figure_SignalExtraction_7090}
\end{figure}

In order to extract the signal, the unbinned data $N(m_{ee}, p_{\rm T})$ is fitted with the model from Eq.~\ref{eq:FitEq}, where the only free parameters are $f_{\rm hadr}$ and $f_{\rm phot}$. The fit is performed in the $p_{\rm T}$ range 0 $< p_{\rm T} <$ 7 GeV/$c$ and the mass range 2 $< m_{ee} <$ 4 GeV/$c^2$. This procedure is performed in the centrality classes \mbox{40--50\%}, \mbox{50--70\%}, and \mbox{70--90\%}, and the raw yield of coherently photoproduced \jpsi, $N_{\jpsi, {\rm coh}}$, is found by integrating the corresponding fitted template function for $p_{\rm T}<$0.2~\GeVc and 2.92$< m_{\ee} < $ 3.16~\GeVmass. 
The quality of the fitting procedure is illustrated in Fig.~\ref{fig:figure_SignalExtraction_7090}. The left panel shows a projection on the invariant mass dimension for candidate pairs with $\pt<0.2$~\GeVc for \PbPb collisions in the 70--90\% centrality range. The right panel shows a projection on the \pt dimension for pairs in the invariant mass interval $2.92< m_{\ee} < 3.16$~\GeVmass and the same centrality range. The two projections of the candidate pair distribution are compared to the projection from the fit model and its various components as described above.
\begin{table}[h!]
  \begin{center}
  \caption{Values of the $\chi^2/\rm{ndf}$ for the invariant mass and \pt projections computed with respect to the fit model projections for the three analyzed centrality intervals.}
    \begin{tabular}{l|c|c|c}
      \hline 
      \textbf{Centrality} &\textbf{40--50\%}& \textbf{50--70\%} & \textbf{70--90\%}\\
       \hline
      Invariant mass projection (full fit range) & 0.7 & 0.8 & 1.4 \\
      
      \pt projection ($\pt < 0.2$~\GeVc) & 1.6 & 1.9 & 1.1 \\
    
      \pt projection (full fit range) & 1.6 & 1.6 & 1.3 \\
     \hline 
    \end{tabular}
    
    \label{tab:table2_chi2}
  \end{center}
\end{table}
The quality of the fit for the three analyzed centrality ranges was estimated quantitatively by calculating the $\chi^2/\rm{ndf}$ for both the invariant mass and \pt projections and is listed in Table~\ref{tab:table2_chi2}.

The doubly-differential cross section of the coherent \jpsi photoproduction, ${\rm d}^2 \sigma /{\rm d}y {\rm d}p_{\rm T}$, for a rapidity interval, $\Delta y$, and a \pt interval, $\Delta \pt$, is computed as:
\begin{equation}
  \frac{{\rm d}^2 \sigma}{{\rm d}y {\rm d}p_{\rm T}} =
  \frac{N_{\jpsi, {\rm coh}}}{(A \times \epsilon)_{\jpsi, {\rm coh}} \times
    BR(\jpsi \rightarrow e^+ e ^-) \times \Delta {\mathit y} \times
    \Delta  {\mathit p_{\rm T}} \times \mathcal{L}}~,
\end{equation}
where $N_{\jpsi, {\rm coh}}$ is the raw yield of coherently photoproduced \jpsi, $(A \times \epsilon)_{\jpsi, {\rm coh}}$ is the average acceptance and efficiency factor for the kinematic window studied, and $\mathcal{L}$ is the integrated luminosity for the analysed data sample as given in Sec.~\ref{sec:experiment}, all referring to the centrality class considered. The obtained computed cross sections correspond to the luminosity integrated in the given centrality interval. The $BR(\jpsi \rightarrow e^+ e ^-)$ is the branching ratio of the \jpsi decay into the dielectron channel~\cite{ParticleDataGroup:2022pth}.  

The acceptance times efficiency correction is the product of the
kinematic acceptance, the tracking and particle identification (PID) efficiency, and the fraction of the signal contained within the invariant mass counting window. With the exception of the PID efficiency, all these factors are obtained based on a MC simulation of coherently photoproduced, transversely polarised \jpsi generated using STARLight and embedded in Pb--Pb collisions  generated using HIJING, as described above.
The \jpsi PID efficiency is estimated using a data-driven method, similar to the one described in Refs.~\cite{ALICE:2021dtt,ALICE:2023gco}, based on a pure sample of electrons using tagged photon conversions.  
The total acceptance times efficiency correction is about 10$\%$ on average,
with a \pt dependence which for the contributing factors listed above is quite mild
in the \mbox{\pt $<$ 0.5 GeV/$c$} range relevant here. However, there is some \pt migration
between generator and detector level due to the \jpsi \pt resolution being of the same order as the typical \pt of the coherent photoproduced \jpsi, and due to the radiative \jpsi decay. The impact of bin migration was checked by applying an unfolding procedure to the measured raw spectrum, as described later in this section, and the differences were found to be negligible. 
 
The systematic uncertainties affecting the measured cross sections originate from uncertainties on the tracking and particle identification efficiencies, signal extraction, collision centrality, luminosity determination, and the branching ratio of the \jpsi into dielectrons. A summary of all the uncertainties is provided in Table~\ref{tab:table1_totalSyst}.
The tracking uncertainty consists of two contributions, one from the ITS-TPC matching and one due to the choice of the track quality criteria, as also detailed in Ref.~\cite{ALICE:2021dtt}.  The former is taken as the difference observed between MC simulations and data for the single-track ITS-TPC matching efficiency, propagated to \jpsi candidate dielectrons, and amounts to about 6.5\%. The latter is estimated by repeating the analysis with variations of the track quality criteria and amounts to approximately 3.3\%. The uncertainty on the PID efficiency is estimated by varying the pion and
proton exclusion cuts, as described in Refs.~\cite{ALICE:2023gco,ALICE:2021dtt}, and resulted in an uncertainty of 2.1\% for the 40--50\% and 3.8\% for the 50--90\% centrality ranges.
The uncertainty on the signal extraction includes contributions from the templates used in the fit model described in Eq.~\ref{eq:FitEq}, fit ranges, and signal counting method. 
The $F_{\rm bkg}$ template was changed by removing the correlated background component and thus keeping only the combinatorial component estimated with event mixing. 
Adopting a rather conservative approach, the $F_{\rm hadr}$ component
is constructed with and without the tuning of the input \pt shape 
on experimental data, and the associated systematic error is taken as $1/\sqrt{12}$ of the difference of the resulting cross sections, corresponding to the ratio of the standard deviation and the full spread of a uniform distribution.
The corresponding uncertainties from the $F_{\rm bkg}$ and $F_{\rm hadr}$ variations range between 0.2--4.1\% and 0.6--1.1\%, respectively. 
The $F_{\rm phot}$ template corresponding to the coherent photoproduction process was varied from its shape expected for UPC, assuming that the $\langle p_{\rm T} \rangle$ of the distribution is inversely proportional to the radius of the target nucleus spectator region, with the radius being determined for each centrality range as $R\sim A_{\rm spec}^{1/3}$, with $A_{\rm spec} = A_{\rm Pb} - \langle N_{\rm part} \rangle /2$. Hence, the template shapes were modified for each centrality interval by shifting each entry in the original STARLight-generated template
from \pt to $r$\pt, with $r = (A_{\rm Pb}/ A_{\rm spec})^{1/3}$,
plus a Gaussian smearing term, centred at zero, to avoid binning effects.  
The \jpsi yields extracted with these modified templates deviated from the default between 0.15 and 3.1\%, depending on centrality, with the deviation growing towards more central collisions as expected.
Variations of the fit ranges in invariant mass and transverse momentum led to systematic uncertainties between 0.35 and 2.2\%, depending on centrality. The default signal counting method, integrating the \jpsi coherent photoproduction template, was changed to counting the entries left after subtracting all the other fit components. The extracted yields changed by 18.5\%, 3.5\%, and 1.3\% for the centrality ranges 40--50\%, 50--70\%, and 70--90\%, respectively.
The uncertainty on the integrated luminosity is 2.5\%  and is determined in the analysis of the van der Meer scan runs as described in Ref.~\cite{ALICE:2022xir}. The uncertainty on the centrality selection is determined by running the analysis with centrality ranges shifted by 1\% in either direction to account for uncertainties on the anchor point used in the centrality definition~\cite{ALICE-PUBLIC-2018-011}. The variations of the extracted yields are compatible with statistical fluctuations, but we conservatively assigned a systematic uncertainty of 1\% for the most peripheral centrality range and 2\% for the rest. For the ($y, p_{\rm T}$) doubly-differential cross sections, the
systematic uncertainties are considered to a large extent correlated over \pt and uniform (global) in the \pt range relevant for coherent \jpsi photoproduction. An exception is the uncertainty due to the fit procedure, which is also strongly correlated over $p_{\rm T}$; however, the size of the uncertainties have point-to-point variations and thus are not considered as global.
The uncertainty on the \jpsi $\rightarrow e^+ e^-$ branching
ratio is quoted as 0.5\%~\cite{ParticleDataGroup:2022pth}. 

Two methodical cross checks were performed for the analysis. For the first one, the fixed weights $w_{{\rm phot},i}$ adopted from Ref.~\cite{ALICE:2021gpt} for the photoproduced components $F_{{\rm phot},i}$, were grouped according to their origin: the weights for the coherent and incoherent charmonium production plus the \mbox{$\gamma \gamma \rightarrow e^+e^-$} continuum were left free in the fit, while keeping the same ratio as in Ref.~\cite{ALICE:2021gpt} between feed-down and direct production for the coherent and incoherent components. This approach gave very similar results to the default method, and no associated systematic error was assigned. The second cross check involved using an unfolding procedure instead of the acceptance times efficiency correction
to obtain the doubly-differential cross sections,
in order to quantify possible effects due to the \jpsi kinematic variation between the generator and detector level. The response matrix was constructed using the MC simulations for coherently photoproduced \jpsi, and the unfolding was performed using both the Bayes~\cite{DAgostini:1994fjx} and Single Value Decomposition~\cite{Hocker:1995kb} methods with different values for the regularisation parameter. The difference relative to the default method was found to be negligible in this case as well, and no systematic uncertainty was assigned.   
\begin{table}[h!]
  \begin{center}
  \caption{Systematic uncertainties on the coherent \jpsi cross section for different centrality intervals. The systematic sources marked with a ${(*)}$ are considered to be correlated over centrality. All the systematic uncertainty sources are considered fully correlated over \pt.}
    \begin{tabular}{l|c|c|c}
      \hline 
      \textbf{Centrality} &\textbf{40--50\%}& \textbf{50--70\%} & \textbf{70--90\%}\\
       \hline
      Tracking (cut variations)$^{*}$ &3.3\% & 3.3\% & 3.3\%\\
      
      Tracking (ITS-TPC matching)$^{*}$ &6.5\% & 6.5\% & 6.5\%\\
    
      Electron identification$^{*}$ &2.1\%& 3.8\% & 3.8\%\\
    
      Fit (correlated bkg) &4.1\%& 0.4\% &0.2\%\\
      
      Fit (hadronic \jpsi) & 0.6\%& 1.3\% & 1.1\%\\
      
      Fit (coherent \jpsi) &3.1\%& 0.7\% &0.15\%\\

      Fit range &2.2\%& 1.0\% & 0.4\%\\

      Signal extraction method  &18.5\%& 3.5\% & 1.3\%\\

      Luminosity$^{*}$ &2.5\%& 2.5\% & 2.5\%\\

      Centrality definition$^{*}$ & 2\% & 2\% & 1\% \\

      Branching ratio$^{*}$ & 0.5\% & 0.5\% & 0.5\% \\
     \hline 
      Total &21.1\%&9.7\% & 8.9\%\\
     \hline 
    \end{tabular}
    
    \label{tab:table1_totalSyst}
  \end{center}
\end{table}

\section{Results}
\label{sec:results}

\begin{figure}[tb]
 \centering
    \includegraphics[width = 0.49\textwidth]{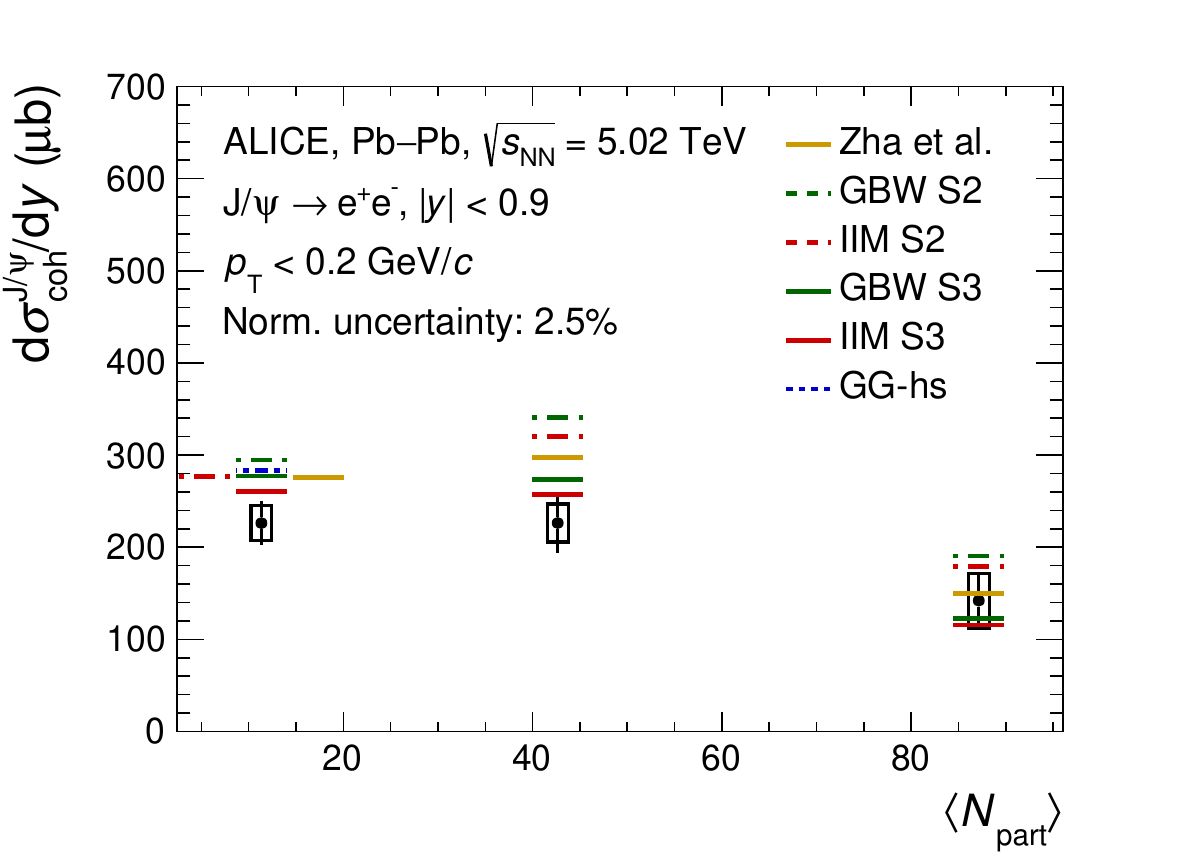}
    \includegraphics[width = 0.49\textwidth]{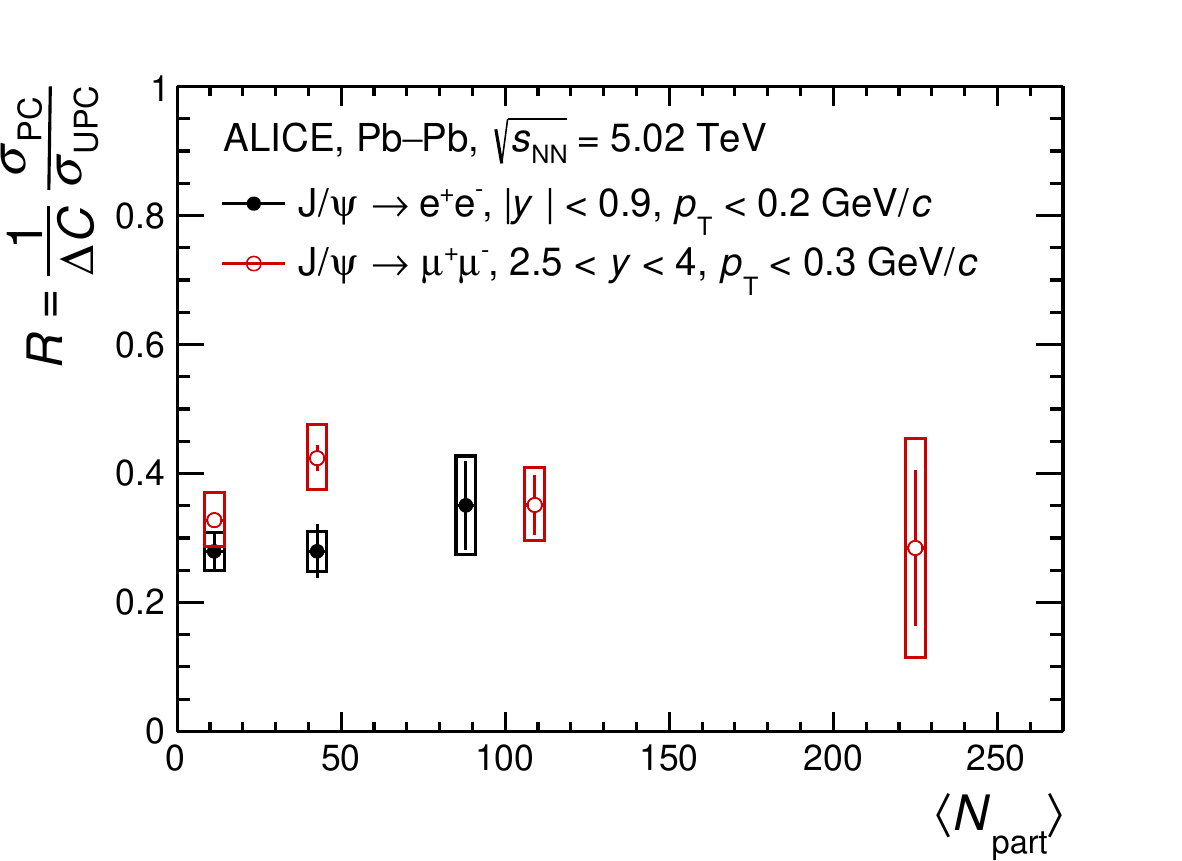}
    \caption{Left: Coherent \jpsi cross section as a function of $\langle \Npart \rangle$ in \PbPb collisions at \fivenn at midrapidity. The statistical uncertainties are shown as bars, while the systematic uncertainties (excluding the global contribution from the beam luminosity) are shown as boxes. It should be noted that the interval 40--50\% is half as wide as the two more peripheral ones, 50--70\% and 70--90\%. Data are compared to calculations from W. Zha {\em et al.}~\cite{Zha:2017jch, Zha:2018jin}, Gay-Ducati {\em et al.}~(GBW S2/S3 and IIM S2/S3)~\cite{GayDucati:2018who}, and Cepila {\em et al.}~(GG-hs)~\cite{Cepila:2017nef}. For the interval 70--90\%, the calculations from W. Zha {\em et al.} and IIM S2 are shifted on the x-axis for visibility. Right: Coherent \jpsi cross section as a function of $\langle \Npart \rangle$ measured at midrapidity (black markers) and forward rapidity (red markers)~\cite{ALICE:2022zso} normalised to the corresponding cross sections measured in the same rapidity ranges in UPC~\cite{ALICE:2019tqa, ALICE:2021gpt} and corrected for the 
    centrality interval width $\Delta C$. }
    \label{fig:figure_CrossSectionvsNpart}
\end{figure}

The left panel of Fig.~\ref{fig:figure_CrossSectionvsNpart} shows the coherent \jpsi photoproduction cross section
${\rm d} \sigma / {\rm d}y$, extracted at
midrapidity ($|y| <$ 0.9), as a function of $\langle N_{\rm part} \rangle$, for the centrality classes 70--90\%, 50--70\%, and 40--50$\%$. 
Systematic uncertainties (excluding the global contribution from the beam luminosity) are depicted as open boxes, while the global centrality-independent systematic uncertainty of 2.5\% is quoted in the legend.
The measured cross sections show a mild centrality dependence, within
uncertainties compatible with no variation within the studied centrality range. It should be noted that, as pointed out in the Sec.~\ref{sec:analysis}, the cross sections are integrated in their respective centrality intervals and that the three data points cover different fractions of the total \PbPb cross section. Namely, the semicentral data point (40--50\%) covers 10\% while the two most peripheral data points each cover 20\% of the total cross section.
The experimental results are compared with a set of theoretical models, as indicated in the legend.    

The calculations are based on the UPC description of vector 
meson photoproduction in terms of a convolution of the photon flux and the photonuclear cross section, but integrating only over the impact parameter range corresponding to the centrality class of interest. The overlapping region in the collision, with hadronic interactions, is considered in some of the calculations by introducing modifications of the photon flux and/or the photonuclear cross section, depending on the model.  
The GG-hs model by Cepila {\em et al.}~\cite{Cepila:2017nef} includes no modifications relative to the UPC picture except the specification of the impact parameter interval. Attempting to describe both coherent and incoherent photoproduction, this model employs an energy-dependent representation of subnucleonic degrees of freedom. This is modelled in terms of hot spots and a colour-dipole proton cross section taking low-$x$ saturation effects into account~\cite{Golec-Biernat:1998zce}, and using the Glauber-Gribov formalism to extrapolate the calculation from nucleonic to nuclear targets. 
In the set of predictions by Gay-Ducati {\em et al.}~\cite{GayDucati:2018who} labelled GBW and IIM, an effective photon flux is defined by integrating over photons reaching the nuclear target and disregarding those hitting the overlap region.
These calculations consider two scenarios, denoted in Fig.~\ref{fig:figure_CrossSectionvsNpart} as S2 and S3, with S2 modifying the photon flux only while S3 in addition restricts the photonuclear cross section, excluding contributions from the hadronically interacting overlap zone. The GBW~\cite{Golec-Biernat:1998zce} and the IIM~\cite{Iancu:2003ge} descriptions employ different treatments of saturation effects in the colour-dipole proton cross section, based on parameterisations of DIS data. 
The approach by Zha {\em et al.}~\cite{Zha:2017jch, Zha:2018jin} takes into account the potential effect on photon and gluon emission by the hadronically interacting overlapping zone, exploring different scenarios regarding the roles of participants and spectators. In addition, these calculations consider the implications of the destructive interference between the photon amplitudes from the two nuclear sources moving in opposite directions.  The estimates from Ref.~\cite{Zha:2018jin} shown in Fig.~\ref{fig:figure_CrossSectionvsNpart} include nuclear shadowing effects and assume an unaltered photon emission from the overlapping region except for interference effects, combined with the disruption of photoproduction in the participant region.

The experimental results are qualitatively well described by the model calculations, although for the two most peripheral intervals, the absolute values of the cross sections are overestimated by all models. 
In particular, the calculations by Gay-Ducati {\em et al.}~\cite{GayDucati:2018who} for scenario S3, which take into account suppression of photoproduction in the overlap zone, are the most compatible with the data with a statistical significance of the differences averaged over the three centrality intervals of less than 2$\sigma$. However, the S2 scenario from Ref.~\cite{GayDucati:2018who} overestimates the measured cross section for all studied centrality intervals and the statistical compatibility with the data is around 3$\sigma$. 
It is worth noting that a modification of the photonuclear cross section might not be the only
mechanism that explains the data. ALICE measurements at forward rapidity~\cite{ALICE:2022zso}, which are qualitatively similar in trend and model comparisons to the results at midrapidity, are well described by
the calculations by Klusek-Gawenda~\cite{Klusek-Gawenda:2015hja}, which only assume inhibition of the photon flux in the overlap region.

The comparison with data at forward rapidity might also provide information on any final-state QGP influence on photoproduced charmonia. Such effects are naively expected to be more pronounced at midrapidity, where a higher medium temperature and energy density are expected~\cite{ALICE:2022wpn}. The \jpsi photoproduction at mid- and forward rapidity are compared after normalising the measurements to the coherent
\jpsi photoproduction cross section measured in UPC in the same rapidity interval in order to compensate for the rapidity dependence of photon flux and photo-nuclear cross section. Uncertainties which are correlated between the results in collisions with overlap and those in UPC are cancelled in the ratio. There are also two small differences in the kinematic coverage of the UPC results compared to those from peripheral or semicentral collisions. Namely, the midrapidity UPC coherent \jpsi yield is measured at $|y| < 0.8$ and the \pt range for the forward-rapidity one is $\pt < 0.25$~\GeVc. This ratio ($\sigma_{\rm PC} / \sigma_{\rm UPC}$), corrected for the different centrality interval widths $\Delta {\rm C}$, is shown in the right panel of Fig.~\ref{fig:figure_CrossSectionvsNpart} as a function of $\langle N_{\rm part} \rangle$, for forward \mbox{(2.5 $< y <$ 4)} and midrapidity \mbox{($|y| <$ 0.9)}. The measurements at forward rapidity in peripheral collisions are taken from Ref.~\cite{ALICE:2022zso}, while the UPC results for the two rapidity intervals are from Refs.~\cite{ALICE:2019tqa, ALICE:2021gpt}. Within uncertainties, the ratios at the two rapidities are statistically compatible over the common centrality range, and show a nearly flat evolution with $\langle N_{\rm part} \rangle$. Notably, the measurements for the most peripheral collisions, 70--90\%, which are also the most precise, are in very good agreement. 
The measured cross sections in this analysis show no hints of QGP effects,
but the centrality range studied is rather limited, extending only
up to $\langle N_{\rm part} \rangle <$ 100, which, combined with the
large experimental uncertainties, may be insufficient to reveal 
any medium-induced suppression. Additional rapidity dependent effects on the photon flux due to the nuclear overlap may complicate this 
naive phenomenological picture, hence further model calculations are required for a deeper insight.
The large increase in luminosity to be collected during the LHC Run 3 and 4 is expected to extend the range of the \jpsi measurement up to the most central collisions and also give access to the centrality dependent photoproduction of the less bound $\psi(2S)$, important for shedding light on these phenomenological aspects.

\begin{figure}[tb]
 \centering
   \includegraphics[width = 0.49\textwidth]{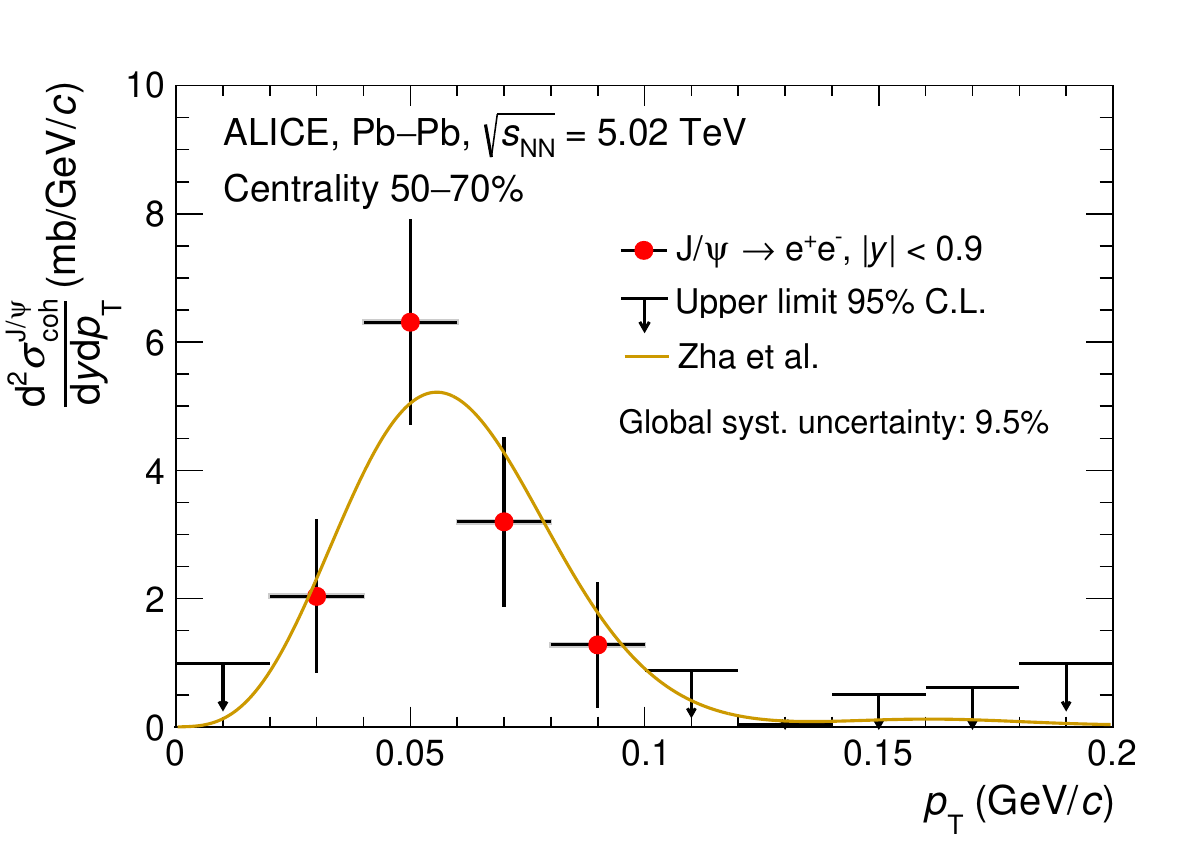}
   \includegraphics[width = 0.49\textwidth]{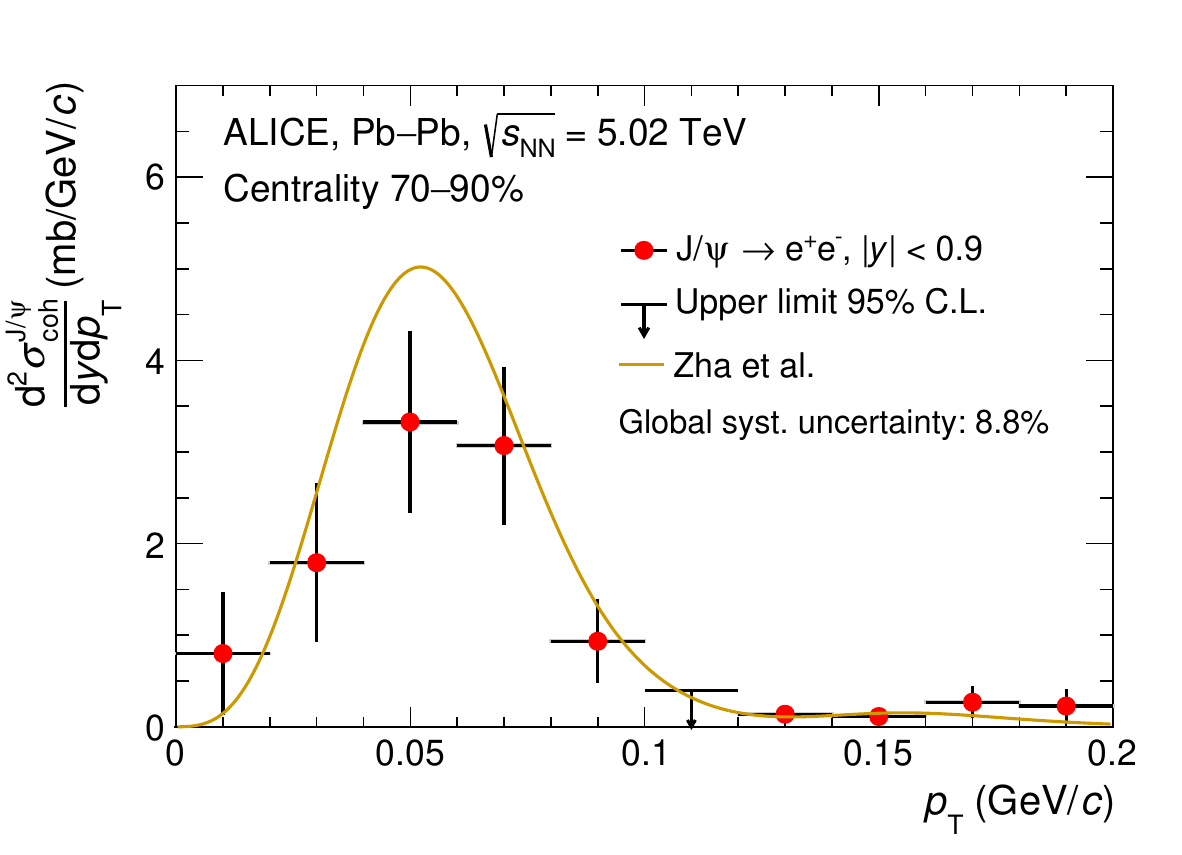}
    \caption{Coherent \jpsi cross section as a function of \pt in \PbPb collisions at \fivenn measured in the 50--70\% and 70--90\% centrality classes. The error bars indicate statistical uncertainties, while the systematic uncertainties, excluding those considered as global (see text for details), are shown as boxes. Data are compared to calculations from W. Zha {\em et al.}~\cite{Zha:2017jch, Zha:2018jin}.}
    \label{fig:figure_CrossSectionvsPt}
\end{figure}

Figure~\ref{fig:figure_CrossSectionvsPt} shows the doubly-differential cross section ${\rm d}^2 \sigma / {\rm d} y {\rm d}p_{\rm T}$ for the rapidity interval $|y|<0.9$ and in the centrality classes 50--70\% (left) and 70--90\% (right), which was obtained by subtracting from the measured distribution $N (\mee, \pt)$ all of the fitted templates with the exception of the one corresponding to the coherently photoproduced \jpsi. The \pt-independent systematic uncertainties are quoted in the legend for the two centralities as global uncertainties, while the point-to-point systematic uncertainties related to the fit of the two-dimensional $(\mee,\pt)$ pair distributions are shown as boxes around the data points and are very small compared to the global ones and the statistical uncertainties.
The theoretical calculations by Zha {\em et al.}, described above~\cite{Zha:2017jch,Zha:2018jin}, in which the nucleons in the overlapping zone do not participate in the photoproduction, are in good agreement with ALICE measurements within the large experimental uncertainties.
This model was also found to reproduce the shape of the LHCb \pt differential
measurements, but appears to underpredict the reported yield~\cite{LHCb:2021hoq}.
In the model, the rapid decline of the cross section towards \pt = 0 is ascribed to the destructive interference between photon amplitudes from the two collision partners, while the bulk of the distribution and the tail towards higher \pt carry information about the target size and spatial distribution.  In particular, a variation
of the \pt differential cross section is expected with centrality.
The observed shapes of the \pt spectra in the 50--70\% and 70--90\%
centrality classes are compatible with the current experimental uncertainties, however, only a very weak centrality dependence is predicted in this centrality range, as seen from the theory calculation shown in Fig.~\ref{fig:figure_CrossSectionvsPt}.

\section{Conclusions}
\label{sec:conclusion}

In summary, this paper reports on the measurement of coherent \jpsi photoproduction cross section as a function of \pt and collision centrality, for peripheral and semicentral \PbPb collisions at \mbox{\snn = 5.02~TeV}, for the first time at midrapidity ($|y| < 0.9$) at LHC energies. 
The reported observations extend and complement earlier measurements published by ALICE and LHCb at forward rapidity, and STAR at midrapidity at a lower collision energy. The \pt-integrated cross section exhibits a mild centrality dependence compatible with no variation over the centrality interval studied and is qualitatively similar to the corresponding observations at forward rapidity. Within the large uncertainties, the centrality-dependent measurements are fairly well described by several sets of theoretical calculations, based on the UPC description of vector meson photoproduction and modified to take into account the nuclear overlap in the collisions, but not incorporating effects from a hot expanding QGP.
The combined ALICE measurements at mid- and forward rapidity 
favour models with inhibition of photon flux only, or inhibition of the photon flux together with a suppression
of charmonium generation in the hadronic overlap region.
Within uncertainties, the measured cross sections show no suppression beyond the present model predictions. The \pt-integrated cross sections measured at mid- and forward rapidity by ALICE, normalised to the cross sections measured in the same rapidity ranges in UPC~\cite{ALICE:2019tqa, ALICE:2021gpt}, are in good agreement within uncertainties. 
The \pt-differential cross sections are well described by calculations
taking into account the interference between the two photon emitters as
well as the effect of strong interactions in the overlap zone.  
Future data-taking campaigns during the LHC Run 3 and Run~4~\cite{ALICE:2012dtf}, with a foreseen increase in statistics of a factor of about 100 at midrapidity, are expected to facilitate photoproduction measurements for central collisions (0--10\%) and precision measurements of $p_{\rm T}$-differential cross sections for non-central collisions (above 10\%). This will help elucidate the influence of the nuclear overlap region on the coherent \jpsi photoproduction, including possible final-state effects from the expanding QGP.


\newenvironment{acknowledgement}{\relax}{\relax}
\begin{acknowledgement}
\section*{Acknowledgements}

The ALICE Collaboration would like to thank all its engineers and technicians for their invaluable contributions to the construction of the experiment and the CERN accelerator teams for the outstanding performance of the LHC complex.
The ALICE Collaboration gratefully acknowledges the resources and support provided by all Grid centres and the Worldwide LHC Computing Grid (WLCG) collaboration.
The ALICE Collaboration acknowledges the following funding agencies for their support in building and running the ALICE detector:
A. I. Alikhanyan National Science Laboratory (Yerevan Physics Institute) Foundation (ANSL), State Committee of Science and World Federation of Scientists (WFS), Armenia;
Austrian Academy of Sciences, Austrian Science Fund (FWF): [M 2467-N36] and Nationalstiftung f\"{u}r Forschung, Technologie und Entwicklung, Austria;
Ministry of Communications and High Technologies, National Nuclear Research Center, Azerbaijan;
Conselho Nacional de Desenvolvimento Cient\'{\i}fico e Tecnol\'{o}gico (CNPq), Financiadora de Estudos e Projetos (Finep), Funda\c{c}\~{a}o de Amparo \`{a} Pesquisa do Estado de S\~{a}o Paulo (FAPESP) and Universidade Federal do Rio Grande do Sul (UFRGS), Brazil;
Bulgarian Ministry of Education and Science, within the National Roadmap for Research Infrastructures 2020-2027 (object CERN), Bulgaria;
Ministry of Education of China (MOEC) , Ministry of Science \& Technology of China (MSTC) and National Natural Science Foundation of China (NSFC), China;
Ministry of Science and Education and Croatian Science Foundation, Croatia;
Centro de Aplicaciones Tecnol\'{o}gicas y Desarrollo Nuclear (CEADEN), Cubaenerg\'{\i}a, Cuba;
Ministry of Education, Youth and Sports of the Czech Republic, Czech Republic;
The Danish Council for Independent Research | Natural Sciences, the VILLUM FONDEN and Danish National Research Foundation (DNRF), Denmark;
Helsinki Institute of Physics (HIP), Finland;
Commissariat \`{a} l'Energie Atomique (CEA) and Institut National de Physique Nucl\'{e}aire et de Physique des Particules (IN2P3) and Centre National de la Recherche Scientifique (CNRS), France;
Bundesministerium f\"{u}r Bildung und Forschung (BMBF) and GSI Helmholtzzentrum f\"{u}r Schwerionenforschung GmbH, Germany;
General Secretariat for Research and Technology, Ministry of Education, Research and Religions, Greece;
National Research, Development and Innovation Office, Hungary;
Department of Atomic Energy Government of India (DAE), Department of Science and Technology, Government of India (DST), University Grants Commission, Government of India (UGC) and Council of Scientific and Industrial Research (CSIR), India;
National Research and Innovation Agency - BRIN, Indonesia;
Istituto Nazionale di Fisica Nucleare (INFN), Italy;
Japanese Ministry of Education, Culture, Sports, Science and Technology (MEXT) and Japan Society for the Promotion of Science (JSPS) KAKENHI, Japan;
Consejo Nacional de Ciencia (CONACYT) y Tecnolog\'{i}a, through Fondo de Cooperaci\'{o}n Internacional en Ciencia y Tecnolog\'{i}a (FONCICYT) and Direcci\'{o}n General de Asuntos del Personal Academico (DGAPA), Mexico;
Nederlandse Organisatie voor Wetenschappelijk Onderzoek (NWO), Netherlands;
The Research Council of Norway, Norway;
Pontificia Universidad Cat\'{o}lica del Per\'{u}, Peru;
Ministry of Science and Higher Education, National Science Centre and WUT ID-UB, Poland;
Korea Institute of Science and Technology Information and National Research Foundation of Korea (NRF), Republic of Korea;
Ministry of Education and Scientific Research, Institute of Atomic Physics, Ministry of Research and Innovation and Institute of Atomic Physics and Universitatea Nationala de Stiinta si Tehnologie Politehnica Bucuresti, Romania;
Ministry of Education, Science, Research and Sport of the Slovak Republic, Slovakia;
National Research Foundation of South Africa, South Africa;
Swedish Research Council (VR) and Knut \& Alice Wallenberg Foundation (KAW), Sweden;
European Organization for Nuclear Research, Switzerland;
Suranaree University of Technology (SUT), National Science and Technology Development Agency (NSTDA) and National Science, Research and Innovation Fund (NSRF via PMU-B B05F650021), Thailand;
Turkish Energy, Nuclear and Mineral Research Agency (TENMAK), Turkey;
National Academy of  Sciences of Ukraine, Ukraine;
Science and Technology Facilities Council (STFC), United Kingdom;
National Science Foundation of the United States of America (NSF) and United States Department of Energy, Office of Nuclear Physics (DOE NP), United States of America.
In addition, individual groups or members have received support from:
Czech Science Foundation (grant no. 23-07499S), Czech Republic;
FORTE project, reg.\ no.\ CZ.02.01.01/00/22\_008/0004632, Czech Republic, co-funded by the European Union, Czech Republic;
European Research Council (grant no. 950692), European Union;
ICSC - Centro Nazionale di Ricerca in High Performance Computing, Big Data and Quantum Computing, European Union - NextGenerationEU;
Academy of Finland (Center of Excellence in Quark Matter) (grant nos. 346327, 346328), Finland.   
\end{acknowledgement}

\bibliographystyle{utphys}   
\bibliography{bibliography}

\providecommand{\href}[2]{#2}\begingroup\raggedright\begin{thebibliography}{10}

\bibitem{Klein:2019qfb}
S.~R. Klein and H.~M\"antysaari, ``{Imaging the nucleus with high-energy
  photons}'', \href{https://doi.org/10.1038/s42254-019-0107-6}{{\em Nature Rev.
  Phys.} {\bfseries 1} (2019) 662--674},
  \href{https://arxiv.org/abs/1910.10858}{{\ttfamily arXiv:1910.10858
  [hep-ex]}}.

\bibitem{Baltz:2007kq}
A.~J. Baltz {\em et~al.}, ``{The Physics of Ultraperipheral Collisions at the
  LHC}'', \href{https://doi.org/10.1016/j.physrep.2007.12.001}{{\em Phys.
  Rept.} {\bfseries 458} (2008) 1--171},
  \href{https://arxiv.org/abs/0706.3356}{{\ttfamily arXiv:0706.3356
  [nucl-ex]}}.

\bibitem{Bertulani:2005ru}
C.~A. Bertulani, S.~R. Klein, and J.~Nystrand, ``{Physics of ultra-peripheral
  nuclear collisions}'',
  \href{https://doi.org/10.1146/annurev.nucl.55.090704.151526}{{\em Ann. Rev.
  Nucl. Part. Sci.} {\bfseries 55} (2005) 271--310},
  \href{https://arxiv.org/abs/nucl-ex/0502005}{{\ttfamily
  arXiv:nucl-ex/0502005}}.

\bibitem{ALICE:2021tyx}
{\bfseries ALICE} Collaboration, S.~Acharya {\em et~al.}, ``{First measurement
  of the $|t|$-dependence of coherent ${\rm J}/\psi$ photonuclear
  production}'', \href{https://doi.org/10.1016/j.physletb.2021.136280}{{\em
  Phys. Lett. B} {\bfseries 817} (2021) 136280},
  \href{https://arxiv.org/abs/2101.04623}{{\ttfamily arXiv:2101.04623
  [nucl-ex]}}.

\bibitem{ALICE:2012yye}
{\bfseries ALICE} Collaboration, B.~Abelev {\em et~al.}, ``{Coherent ${\rm
  J}/\psi$ photoproduction in ultra-peripheral Pb-Pb collisions at
  $\sqrt{s_{NN}} = 2.76$ TeV}'',
  \href{https://doi.org/10.1016/j.physletb.2012.11.059}{{\em Phys. Lett. B}
  {\bfseries 718} (2013) 1273--1283},
  \href{https://arxiv.org/abs/1209.3715}{{\ttfamily arXiv:1209.3715
  [nucl-ex]}}.

\bibitem{ALICE:2013wjo}
{\bfseries ALICE} Collaboration, E.~Abbas {\em et~al.}, ``{Charmonium and
  $e^+e^-$ pair photoproduction at mid-rapidity in ultra-peripheral Pb-Pb
  collisions at $\sqrt{s_{\rm NN}}$=2.76 TeV}'',
  \href{https://doi.org/10.1140/epjc/s10052-013-2617-1}{{\em Eur. Phys. J. C}
  {\bfseries 73} (2013) 2617},
  \href{https://arxiv.org/abs/1305.1467}{{\ttfamily arXiv:1305.1467
  [nucl-ex]}}.

\bibitem{CMS:2016itn}
{\bfseries CMS} Collaboration, V.~Khachatryan {\em et~al.}, ``{Coherent ${\rm
  J}/\psi$ photoproduction in ultra-peripheral PbPb collisions at $\sqrt
  {s_{NN}} =$ 2.76 TeV with the CMS experiment}'',
  \href{https://doi.org/10.1016/j.physletb.2017.07.001}{{\em Phys. Lett. B}
  {\bfseries 772} (2017) 489--511},
  \href{https://arxiv.org/abs/1605.06966}{{\ttfamily arXiv:1605.06966
  [nucl-ex]}}.

\bibitem{ALICE:2019tqa}
{\bfseries ALICE} Collaboration, S.~Acharya {\em et~al.}, ``{Coherent ${\rm
  J}/\psi$ photoproduction at forward rapidity in ultra-peripheral Pb-Pb
  collisions at $\sqrt{s_{\rm{NN}}}=5.02$ TeV}'',
  \href{https://doi.org/10.1016/j.physletb.2019.134926}{{\em Phys. Lett. B}
  {\bfseries 798} (2019) 134926},
  \href{https://arxiv.org/abs/1904.06272}{{\ttfamily arXiv:1904.06272
  [nucl-ex]}}.

\bibitem{ALICE:2021gpt}
{\bfseries ALICE} Collaboration, S.~Acharya {\em et~al.}, ``{Coherent ${\rm
  J}/\psi$ and $\psi'$ photoproduction at midrapidity in ultra-peripheral Pb-Pb
  collisions at $\sqrt{s_{\mathrm{NN}}}~=~5.02$ TeV}'',
  \href{https://doi.org/10.1140/epjc/s10052-021-09437-6}{{\em Eur. Phys. J. C}
  {\bfseries 81} (2021) 712},
  \href{https://arxiv.org/abs/2101.04577}{{\ttfamily arXiv:2101.04577
  [nucl-ex]}}.

\bibitem{LHCb:2021bfl}
{\bfseries LHCb} Collaboration, R.~Aaij {\em et~al.}, ``{Study of coherent
  $J/\psi$ production in lead-lead collisions at $
  \sqrt{{\mathrm{s}}_{\mathrm{NN}}} $ = 5 TeV}'',
  \href{https://doi.org/10.1007/JHEP07(2022)117}{{\em JHEP} {\bfseries 07}
  (2022) 117}, \href{https://arxiv.org/abs/2107.03223}{{\ttfamily
  arXiv:2107.03223 [hep-ex]}}.

\bibitem{ALICE:2023jgu}
{\bfseries ALICE} Collaboration, S.~Acharya {\em et~al.}, ``{Energy dependence
  of coherent photonuclear production of J/\ensuremath{\psi} mesons in
  ultra-peripheral Pb-Pb collisions at $ \sqrt{{\textrm{s}}_{\textrm{NN}}} $ =
  5.02 TeV}'', \href{https://doi.org/10.1007/JHEP10(2023)119}{{\em JHEP}
  {\bfseries 10} (2023) 119},
  \href{https://arxiv.org/abs/2305.19060}{{\ttfamily arXiv:2305.19060
  [nucl-ex]}}.

\bibitem{CMS:2023snh}
{\bfseries CMS} Collaboration, A.~Tumasyan {\em et~al.}, ``{Probing Small
  Bjorken-x Nuclear Gluonic Structure via Coherent J/\ensuremath{\psi}
  Photoproduction in Ultraperipheral Pb-Pb Collisions at \snn=5.02\,\,TeV}'',
  \href{https://doi.org/10.1103/PhysRevLett.131.262301}{{\em Phys. Rev. Lett.}
  {\bfseries 131} (2023) 262301},
  \href{https://arxiv.org/abs/2303.16984}{{\ttfamily arXiv:2303.16984
  [nucl-ex]}}.

\bibitem{STAR:2023nos}
{\bfseries STAR} Collaboration, M.~I. Abdulhamid {\em et~al.}, ``{Observation
  of Strong Nuclear Suppression in Exclusive J/\ensuremath{\psi}
  Photoproduction in Au+Au Ultraperipheral Collisions at RHIC}'',
  \href{https://doi.org/10.1103/PhysRevLett.133.052301}{{\em Phys. Rev. Lett.}
  {\bfseries 133} (2024) 052301},
  \href{https://arxiv.org/abs/2311.13637}{{\ttfamily arXiv:2311.13637
  [nucl-ex]}}.

\bibitem{ALICE:2015mzu}
{\bfseries ALICE} Collaboration, J.~Adam {\em et~al.}, ``{Measurement of an
  excess in the yield of ${\rm J}/\psi$ at very low $p_{\rm T}$ in Pb-Pb
  collisions at $\sqrt{s_{\rm NN}}$ = 2.76 TeV}'',
  \href{https://doi.org/10.1103/PhysRevLett.116.222301}{{\em Phys. Rev. Lett.}
  {\bfseries 116} (2016) 222301},
  \href{https://arxiv.org/abs/1509.08802}{{\ttfamily arXiv:1509.08802
  [nucl-ex]}}.

\bibitem{STAR:2019yox}
{\bfseries STAR} Collaboration, J.~Adam {\em et~al.}, ``{Observation of excess
  ${\rm J}/\psi$ yield at very low transverse momenta in Au+Au collisions at
  $\sqrt{s_{\rm{NN}}} =$ 200 GeV and U+U collisions at $\sqrt{s_{\rm{NN}}} =$
  193 GeV}'', \href{https://doi.org/10.1103/PhysRevLett.123.132302}{{\em Phys.
  Rev. Lett.} {\bfseries 123} (2019) 132302},
  \href{https://arxiv.org/abs/1904.11658}{{\ttfamily arXiv:1904.11658
  [hep-ex]}}.

\bibitem{LHCb:2021hoq}
{\bfseries LHCb} Collaboration, R.~Aaij {\em et~al.}, ``{${\rm J}/\psi$
  photoproduction in Pb-Pb peripheral collisions at $\sqrt {s_{NN}}$= 5 TeV}'',
  \href{https://doi.org/10.1103/PhysRevC.105.L032201}{{\em Phys. Rev. C}
  {\bfseries 105} (2022) L032201},
  \href{https://arxiv.org/abs/2108.02681}{{\ttfamily arXiv:2108.02681
  [hep-ex]}}.

\bibitem{ALICE:2022zso}
{\bfseries ALICE} Collaboration, S.~Acharya {\em et~al.}, ``{Photoproduction of
  low-$p_{\rm T}$ J/$\psi$ from peripheral to central Pb$-$Pb collisions at
  5.02 TeV}'', \href{https://doi.org/10.1016/j.physletb.2022.137467}{{\em Phys.
  Lett. B} {\bfseries 846} (2023) 137467},
  \href{https://arxiv.org/abs/2204.10684}{{\ttfamily arXiv:2204.10684
  [nucl-ex]}}.

\bibitem{GayDucati:2018who}
M.~B. Gay~Ducati and S.~Martins, ``{Heavy meson photoproduction in peripheral
  AA collisions}'', \href{https://doi.org/10.1103/PhysRevD.97.116013}{{\em
  Phys. Rev. D} {\bfseries 97} (2018) 116013},
  \href{https://arxiv.org/abs/1804.09836}{{\ttfamily arXiv:1804.09836
  [hep-ph]}}.

\bibitem{Cepila:2017nef}
J.~Cepila, J.~G. Contreras, and M.~Krelina, ``{Coherent and incoherent
  $\mathrm{J/}\psi$ photonuclear production in an energy-dependent hot-spot
  model}'', \href{https://doi.org/10.1103/PhysRevC.97.024901}{{\em Phys. Rev.
  C} {\bfseries 97} (2018) 024901},
  \href{https://arxiv.org/abs/1711.01855}{{\ttfamily arXiv:1711.01855
  [hep-ph]}}.

\bibitem{Zha:2017jch}
W.~Zha, S.~R. Klein, R.~Ma, L.~Ruan, T.~Todoroki, Z.~Tang, Z.~Xu, C.~Yang,
  Q.~Yang, and S.~Yang, ``{Coherent ${\rm J}/\psi$ photoproduction in hadronic
  heavy-ion collisions}'',
  \href{https://doi.org/10.1103/PhysRevC.97.044910}{{\em Phys. Rev. C}
  {\bfseries 97} (2018) 044910},
  \href{https://arxiv.org/abs/1705.01460}{{\ttfamily arXiv:1705.01460
  [nucl-th]}}.

\bibitem{Zha:2018jin}
W.~Zha, L.~Ruan, Z.~Tang, Z.~Xu, and S.~Yang, ``{Double-slit experiment at
  fermi scale: coherent photoproduction in heavy-ion collisions}'',
  \href{https://doi.org/10.1103/PhysRevC.99.061901}{{\em Phys. Rev. C}
  {\bfseries 99} (2019) 061901},
  \href{https://arxiv.org/abs/1810.10694}{{\ttfamily arXiv:1810.10694
  [hep-ph]}}.

\bibitem{Klusek-Gawenda:2015hja}
M.~K\l{}usek-Gawenda and A.~Szczurek, ``{Photoproduction of ${\rm J}/\psi$
  mesons in peripheral and semicentral heavy ion collisions}'',
  \href{https://doi.org/10.1103/PhysRevC.93.044912}{{\em Phys. Rev. C}
  {\bfseries 93} (2016) 044912},
  \href{https://arxiv.org/abs/1509.03173}{{\ttfamily arXiv:1509.03173
  [nucl-th]}}.

\bibitem{Shi:2017qep}
W.~Shi, W.~Zha, and B.~Chen, ``{Charmonium Coherent Photoproduction and
  Hadroproduction with Effects of Quark Gluon Plasma}'',
  \href{https://doi.org/10.1016/j.physletb.2017.12.055}{{\em Phys. Lett. B}
  {\bfseries 777} (2018) 399--405},
  \href{https://arxiv.org/abs/1710.00332}{{\ttfamily arXiv:1710.00332
  [nucl-th]}}.

\bibitem{Jenkovszky:2022wcw}
L.~Jenkovszky, V.~Libov, and M.~V.~T. Machado, ``{Regge phenomenology and
  coherent photoproduction of \jpsi in peripheral heavy ion collisions}'',
  \href{https://doi.org/10.1016/j.physletb.2022.137004}{{\em Phys. Lett. B}
  {\bfseries 827} (2022) 137004},
  \href{https://arxiv.org/abs/2202.02162}{{\ttfamily arXiv:2202.02162
  [hep-ph]}}.

\bibitem{Aamodt:2008zz}
{\bfseries ALICE} Collaboration, K.~Aamodt {\em et~al.}, ``{The ALICE
  experiment at the CERN LHC}'',
\href{https://doi.org/10.1088/1748-0221/3/08/S08002}{{\em JINST} {\bfseries 3}
  (2008) S08002}.

\bibitem{ALICE:2014sbx}
{\bfseries ALICE} Collaboration, B.~B. Abelev {\em et~al.}, ``{Performance of
  the ALICE Experiment at the CERN LHC}'',
  \href{https://doi.org/10.1142/S0217751X14300440}{{\em Int. J. Mod. Phys. A}
  {\bfseries 29} (2014) 1430044},
  \href{https://arxiv.org/abs/1402.4476}{{\ttfamily arXiv:1402.4476
  [nucl-ex]}}.

\bibitem{ALICE:2010tia}
{\bfseries ALICE} Collaboration, K.~Aamodt {\em et~al.}, ``{Alignment of the
  ALICE Inner Tracking System with cosmic-ray tracks}'',
  \href{https://doi.org/10.1088/1748-0221/5/03/P03003}{{\em JINST} {\bfseries
  5} (2010) P03003}, \href{https://arxiv.org/abs/1001.0502}{{\ttfamily
  arXiv:1001.0502 [physics.ins-det]}}.

\bibitem{Alme:2010ke}
J.~Alme {\em et~al.}, ``{The ALICE TPC, a large 3-dimensional tracking device
  with fast readout for ultra-high multiplicity events}'',
  \href{https://doi.org/10.1016/j.nima.2010.04.042}{{\em Nucl. Instrum. Meth.}
  {\bfseries A622} (2010) 316--367},
\href{https://arxiv.org/abs/1001.1950}{{\ttfamily arXiv:1001.1950
  [physics.ins-det]}}.

\bibitem{ALICE:2013axi}
{\bfseries ALICE} Collaboration, E.~Abbas {\em et~al.}, ``{Performance of the
  ALICE VZERO system}'',
  \href{https://doi.org/10.1088/1748-0221/8/10/P10016}{{\em JINST} {\bfseries
  8} (2013) P10016}, \href{https://arxiv.org/abs/1306.3130}{{\ttfamily
  arXiv:1306.3130 [nucl-ex]}}.

\bibitem{ALICE:2013hur}
{\bfseries ALICE} Collaboration, B.~Abelev {\em et~al.}, ``{Centrality
  determination of Pb-Pb collisions at $\sqrt{s_{NN}}$ = 2.76 TeV with
  ALICE}'', \href{https://doi.org/10.1103/PhysRevC.88.044909}{{\em Phys. Rev.
  C} {\bfseries 88} (2013) 044909},
  \href{https://arxiv.org/abs/1301.4361}{{\ttfamily arXiv:1301.4361
  [nucl-ex]}}.

\bibitem{Arnaldi:1999zz}
R.~Arnaldi {\em et~al.}, ``{The Zero degree calorimeters for the ALICE
  experiment}'', \href{https://doi.org/10.1016/j.nima.2008.04.009}{{\em Nucl.
  Instrum. Meth. A} {\bfseries 581} (2007) 397--401}. [Erratum:
  Nucl.Instrum.Meth.A 604, 765 (2009)].

\bibitem{ALICE-PUBLIC-2021-001}
{\bfseries ALICE} Collaboration, S.~Acharya {\em et~al.}, ``{ALICE luminosity
  determination for Pb$-$Pb collisions at $\mathbf{\sqrt{s_{\rm NN}}=5.02}$
  TeV}'',. \url{https://cds.cern.ch/record/2749127}.

\bibitem{Blum:2008nqe}
W.~Blum, L.~Rolandi, and W.~Riegler,
  \href{https://doi.org/10.1007/978-3-540-76684-1}{{\em {Particle detection
  with drift chambers}}}.
\newblock Particle Acceleration and Detection. Springer, 2008.

\bibitem{ALICE:2019pid}
{\bfseries ALICE} Collaboration, S.~Acharya {\em et~al.}, ``{Inclusive ${\rm
  J}/$\ensuremath{\psi} production at mid-rapidity in pp collisions at $
  \sqrt{s} $ = 5.02 TeV}'', \href{https://doi.org/10.1007/JHEP10(2019)084}{{\em
  JHEP} {\bfseries 10} (2019) 084},
  \href{https://arxiv.org/abs/1905.07211}{{\ttfamily arXiv:1905.07211
  [nucl-ex]}}.

\bibitem{ALICE:2021dtt}
{\bfseries ALICE} Collaboration, S.~Acharya {\em et~al.}, ``{Inclusive $\text
  {J}/\psi $ production at midrapidity in pp collisions at $\sqrt{s} = 13$
  TeV}'', \href{https://doi.org/10.1140/epjc/s10052-021-09873-4}{{\em Eur.
  Phys. J. C} {\bfseries 81} (2021) 1121},
  \href{https://arxiv.org/abs/2108.01906}{{\ttfamily arXiv:2108.01906
  [nucl-ex]}}.

\bibitem{ALICE:2019nrq}
{\bfseries ALICE} Collaboration, S.~Acharya {\em et~al.}, ``{Centrality and
  transverse momentum dependence of inclusive J/\ensuremath{\psi} production at
  midrapidity in Pb\textendash{}Pb collisions at \snn=5.02 TeV}'',
  \href{https://doi.org/10.1016/j.physletb.2020.135434}{{\em Phys. Lett. B}
  {\bfseries 805} (2020) 135434},
  \href{https://arxiv.org/abs/1910.14404}{{\ttfamily arXiv:1910.14404
  [nucl-ex]}}.

\bibitem{ALICE:2023gco}
{\bfseries ALICE} Collaboration, S.~Acharya {\em et~al.}, ``{Measurements of
  inclusive \jpsi production at midrapidity and forward rapidity in \PbPb
  collisions at \snn = 5.02 TeV}'',
  \href{https://doi.org/10.1016/j.physletb.2024.138451}{{\em Phys. Lett. B}
  {\bfseries 849} (2024) 138451},
  \href{https://arxiv.org/abs/2303.13361}{{\ttfamily arXiv:2303.13361
  [nucl-ex]}}.

\bibitem{Wang:1991hta}
X.-N. Wang and M.~Gyulassy, ``{HIJING: A Monte Carlo model for multiple jet
  production in $\mathrm{pp}$, $\mathrm{pA}$, and $\mathrm{AA}$ collisions}'',
  \href{https://doi.org/10.1103/PhysRevD.44.3501}{{\em Phys. Rev. D} {\bfseries
  44} (1991) 3501--3516}.

\bibitem{Golonka:2005pn}
P.~Golonka and Z.~Was, ``{PHOTOS Monte Carlo: A Precision tool for QED
  corrections in $Z$ and $W$ decays}'',
  \href{https://doi.org/10.1140/epjc/s2005-02396-4}{{\em Eur. Phys. J. C}
  {\bfseries 45} (2006) 97--107},
  \href{https://arxiv.org/abs/hep-ph/0506026}{{\ttfamily
  arXiv:hep-ph/0506026}}.

\bibitem{Brun:1119728}
R.~Brun, F.~Bruyant, M.~Maire, A.~C. McPherson, and P.~Zanarini, {\em {GEANT 3:
  User's guide Geant 3.10, Geant 3.11; rev. version}}.
\newblock CERN, Geneva, 1987.
\newblock \url{https://cds.cern.ch/record/1119728}.

\bibitem{Klein:2016yzr}
S.~R. Klein, J.~Nystrand, J.~Seger, Y.~Gorbunov, and J.~Butterworth,
  ``{STARlight: A Monte Carlo simulation program for ultra-peripheral
  collisions of relativistic ions}'',
  \href{https://doi.org/10.1016/j.cpc.2016.10.016}{{\em Comput. Phys. Commun.}
  {\bfseries 212} (2017) 258--268},
  \href{https://arxiv.org/abs/1607.03838}{{\ttfamily arXiv:1607.03838
  [hep-ph]}}.

\bibitem{ALICE:2023svb}
{\bfseries ALICE} Collaboration, S.~Acharya {\em et~al.}, ``{First polarisation
  measurement of coherently photoproduced J/$\psi$ in ultra-peripheral Pb$-$Pb
  collisions at $\sqrt{s_{\rm NN}}$ = 5.02 TeV}'',
  \href{https://arxiv.org/abs/2304.10928}{{\ttfamily arXiv:2304.10928
  [nucl-ex]}}.

\bibitem{H1:2013okq}
{\bfseries H1} Collaboration, C.~Alexa {\em et~al.}, ``{Elastic and
  Proton-Dissociative Photoproduction of ${\rm J/}$\ensuremath{\psi} Mesons at
  HERA}'', \href{https://doi.org/10.1140/epjc/s10052-013-2466-y}{{\em Eur.
  Phys. J. C} {\bfseries 73} (2013) 2466},
  \href{https://arxiv.org/abs/1304.5162}{{\ttfamily arXiv:1304.5162 [hep-ex]}}.

\bibitem{ParticleDataGroup:2022pth}
{\bfseries Particle Data Group} Collaboration, S.~Navas {\em et~al.}, ``{Review
  of particle physics}'',
  \href{https://doi.org/10.1103/PhysRevD.110.030001}{{\em Phys. Rev. D}
  {\bfseries 110} (2024) 030001}.

\bibitem{ALICE:2022xir}
{\bfseries ALICE} Collaboration, S.~Acharya {\em et~al.}, ``{ALICE luminosity
  determination for Pb$-$Pb collisions at $\sqrt{s_{\mathrm{NN}}} = 5.02$
  TeV}'', \href{https://doi.org/10.1088/1748-0221/19/02/P02039}{{\em JINST}
  {\bfseries 19} (2024) P02039},
  \href{https://arxiv.org/abs/2204.10148}{{\ttfamily arXiv:2204.10148
  [nucl-ex]}}.

\bibitem{ALICE-PUBLIC-2018-011}
{\bfseries ALICE} Collaboration, S.~Acharya {\em et~al.}, ``{Centrality
  determination in heavy ion collisions}'',.
  \url{https://cds.cern.ch/record/2636623}. ALICE-PUBLIC-2018-011.

\bibitem{DAgostini:1994fjx}
G.~D'Agostini, ``{A Multidimensional unfolding method based on Bayes'
  theorem}'', \href{https://doi.org/10.1016/0168-9002(95)00274-X}{{\em Nucl.
  Instrum. Meth. A} {\bfseries 362} (1995) 487--498}.

\bibitem{Hocker:1995kb}
A.~Hocker and V.~Kartvelishvili, ``{SVD approach to data unfolding}'',
  \href{https://doi.org/10.1016/0168-9002(95)01478-0}{{\em Nucl. Instrum. Meth.
  A} {\bfseries 372} (1996) 469--481},
  \href{https://arxiv.org/abs/hep-ph/9509307}{{\ttfamily
  arXiv:hep-ph/9509307}}.

\bibitem{Golec-Biernat:1998zce}
K.~J. Golec-Biernat and M.~Wusthoff, ``{Saturation effects in deep inelastic
  scattering at low ${Q}^{2}$ and its implications on diffraction}'',
  \href{https://doi.org/10.1103/PhysRevD.59.014017}{{\em Phys. Rev. D}
  {\bfseries 59} (1998) 014017},
  \href{https://arxiv.org/abs/hep-ph/9807513}{{\ttfamily
  arXiv:hep-ph/9807513}}.

\bibitem{Iancu:2003ge}
E.~Iancu, K.~Itakura, and S.~Munier, ``{Saturation and BFKL dynamics in the
  HERA data at small x}'',
  \href{https://doi.org/10.1016/j.physletb.2004.02.040}{{\em Phys. Lett. B}
  {\bfseries 590} (2004) 199--208},
  \href{https://arxiv.org/abs/hep-ph/0310338}{{\ttfamily
  arXiv:hep-ph/0310338}}.

\bibitem{ALICE:2022wpn}
{\bfseries ALICE} Collaboration, S.~Acharya {\em et~al.}, ``{The ALICE
  experiment: a journey through QCD}'',
  \href{https://doi.org/10.1140/epjc/s10052-024-12935-y}{{\em Eur. Phys. J. C}
  {\bfseries 84} (2024) 813},
  \href{https://arxiv.org/abs/2211.04384}{{\ttfamily arXiv:2211.04384
  [nucl-ex]}}.

\bibitem{ALICE:2012dtf}
{\bfseries ALICE} Collaboration, B.~Abelev {\em et~al.}, ``{Upgrade of the
  ALICE Experiment: Letter Of Intent}'',
  \href{https://doi.org/10.1088/0954-3899/41/8/087001}{{\em J. Phys. G}
  {\bfseries 41} (2014) 087001}.

\end{thebibliography}\endgroup

\newpage
\appendix

%
%

\section{The ALICE Collaboration}
\label{app:collab}
\begin{flushleft} 
\small

S.~Acharya\,\orcidlink{0000-0002-9213-5329}\,$^{\rm 127}$, 
A.~Agarwal$^{\rm 135}$, 
G.~Aglieri Rinella\,\orcidlink{0000-0002-9611-3696}\,$^{\rm 32}$, 
L.~Aglietta\,\orcidlink{0009-0003-0763-6802}\,$^{\rm 24}$, 
M.~Agnello\,\orcidlink{0000-0002-0760-5075}\,$^{\rm 29}$, 
N.~Agrawal\,\orcidlink{0000-0003-0348-9836}\,$^{\rm 25}$, 
Z.~Ahammed\,\orcidlink{0000-0001-5241-7412}\,$^{\rm 135}$, 
S.~Ahmad\,\orcidlink{0000-0003-0497-5705}\,$^{\rm 15}$, 
S.U.~Ahn\,\orcidlink{0000-0001-8847-489X}\,$^{\rm 71}$, 
I.~Ahuja\,\orcidlink{0000-0002-4417-1392}\,$^{\rm 37}$
% ...

\section*{Affiliation Notes}

$^{\rm I}$ Deceased\\
$^{\rm II}$ Also at: Max-Planck-Institut fur Physik, Munich, Germany\\
$^{\rm III}$ Also at: Italian National Agency for New Technologies, Energy and Sustainable Economic Development (ENEA), Bologna, Italy\\
$^{\rm IV}$ Also at: Dipartimento DET del Politecnico di Torino, Turin, Italy\\
$^{\rm V}$ Also at: Yildiz Technical University, Istanbul, T\"{u}rkiye\\
$^{\rm VI}$ Also at: Department of Applied Physics, Aligarh Muslim University, Aligarh, India\\
$^{\rm VII}$ Also at: Institute of Theoretical Physics, University of Wroclaw, Poland\\
$^{\rm VIII}$ Also at: An institution covered by a cooperation agreement with CERN\\

\section*{Collaboration Institutes}

$^{1}$ A.I. Alikhanyan National Science Laboratory (Yerevan Physics Institute) Foundation, Yerevan, Armenia\\
$^{2}$ AGH University of Krakow, Cracow, Poland\\
$^{3}$ Bogolyubov Institute for Theoretical Physics, National Academy of Sciences of Ukraine, Kiev, Ukraine\\
$^{4}$ Bose Institute, Department of Physics  and Centre for Astroparticle Physics and Space Science (CAPSS), Kolkata, India\\
$^{5}$ California Polytechnic State University, San Luis Obispo, California, United States\\
$^{6}$ Central China Normal University, Wuhan, China\\
$^{7}$ Centro de Aplicaciones Tecnol\'{o}gicas y Desarrollo Nuclear (CEADEN), Havana, Cuba\\
$^{8}$ Centro de Investigaci\'{o}n y de Estudios Avanzados (CINVESTAV), Mexico City and M\'{e}rida, Mexico\\
$^{9}$ Chicago State University, Chicago, Illinois, United States\\
$^{10}$ China Institute of Atomic Energy, Beijing, China\\
$^{11}$ China University of Geosciences, Wuhan, China\\
$^{12}$ Chungbuk National University, Cheongju, Republic of Korea\\
$^{13}$ Comenius University Bratislava, Faculty of Mathematics, Physics and Informatics, Bratislava, Slovak Republic\\
$^{14}$ Creighton University, Omaha, Nebraska, United States\\
$^{15}$ Department of Physics, Aligarh Muslim University, Aligarh, India\\
$^{16}$ Department of Physics, Pusan National University, Pusan, Republic of Korea\\
$^{17}$ Department of Physics, Sejong University, Seoul, Republic of Korea\\
$^{18}$ Department of Physics, University of California, Berkeley, California, United States\\
$^{19}$ Department of Physics, University of Oslo, Oslo, Norway\\
$^{20}$ Department of Physics and Technology, University of Bergen, Bergen, Norway\\
$^{21}$ Dipartimento di Fisica, Universit\`{a} di Pavia, Pavia, Italy\\
$^{22}$ Dipartimento di Fisica dell'Universit\`{a} and Sezione INFN, Cagliari, Italy\\
$^{23}$ Dipartimento di Fisica dell'Universit\`{a} and Sezione INFN, Trieste, Italy\\
$^{24}$ Dipartimento di Fisica dell'Universit\`{a} and Sezione INFN, Turin, Italy\\
$^{25}$ Dipartimento di Fisica e Astronomia dell'Universit\`{a} and Sezione INFN, Bologna, Italy\\
$^{26}$ Dipartimento di Fisica e Astronomia dell'Universit\`{a} and Sezione INFN, Catania, Italy\\
$^{27}$ Dipartimento di Fisica e Astronomia dell'Universit\`{a} and Sezione INFN, Padova, Italy\\
$^{28}$ Dipartimento di Fisica `E.R.~Caianiello' dell'Universit\`{a} and Gruppo Collegato INFN, Salerno, Italy\\
$^{29}$ Dipartimento DISAT del Politecnico and Sezione INFN, Turin, Italy\\
$^{30}$ Dipartimento di Scienze MIFT, Universit\`{a} di Messina, Messina, Italy\\
$^{31}$ Dipartimento Interateneo di Fisica `M.~Merlin' and Sezione INFN, Bari, Italy\\
$^{32}$ European Organization for Nuclear Research (CERN), Geneva, Switzerland\\
$^{33}$ Faculty of Electrical Engineering, Mechanical Engineering and Naval Architecture, University of Split, Split, Croatia\\
$^{34}$ Faculty of Engineering and Science, Western Norway University of Applied Sciences, Bergen, Norway\\
$^{35}$ Faculty of Nuclear Sciences and Physical Engineering, Czech Technical University in Prague, Prague, Czech Republic\\
$^{36}$ Faculty of Physics, Sofia University, Sofia, Bulgaria\\
$^{37}$ Faculty of Science, P.J.~\v{S}af\'{a}rik University, Ko\v{s}ice, Slovak Republic\\
$^{38}$ Frankfurt Institute for Advanced Studies, Johann Wolfgang Goethe-Universit\"{a}t Frankfurt, Frankfurt, Germany\\
$^{39}$ Fudan University, Shanghai, China\\
$^{40}$ Gangneung-Wonju National University, Gangneung, Republic of Korea\\
$^{41}$ Gauhati University, Department of Physics, Guwahati, India\\
$^{42}$ Helmholtz-Institut f\"{u}r Strahlen- und Kernphysik, Rheinische Friedrich-Wilhelms-Universit\"{a}t Bonn, Bonn, Germany\\
$^{43}$ Helsinki Institute of Physics (HIP), Helsinki, Finland\\
$^{44}$ High Energy Physics Group,  Universidad Aut\'{o}noma de Puebla, Puebla, Mexico\\
$^{45}$ Horia Hulubei National Institute of Physics and Nuclear Engineering, Bucharest, Romania\\
$^{46}$ HUN-REN Wigner Research Centre for Physics, Budapest, Hungary\\
$^{47}$ Indian Institute of Technology Bombay (IIT), Mumbai, India\\
$^{48}$ Indian Institute of Technology Indore, Indore, India\\
$^{49}$ INFN, Laboratori Nazionali di Frascati, Frascati, Italy\\
$^{50}$ INFN, Sezione di Bari, Bari, Italy\\
$^{51}$ INFN, Sezione di Bologna, Bologna, Italy\\
$^{52}$ INFN, Sezione di Cagliari, Cagliari, Italy\\
$^{53}$ INFN, Sezione di Catania, Catania, Italy\\
$^{54}$ INFN, Sezione di Padova, Padova, Italy\\
$^{55}$ INFN, Sezione di Pavia, Pavia, Italy\\
$^{56}$ INFN, Sezione di Torino, Turin, Italy\\
$^{57}$ INFN, Sezione di Trieste, Trieste, Italy\\
$^{58}$ Inha University, Incheon, Republic of Korea\\
$^{59}$ Institute for Gravitational and Subatomic Physics (GRASP), Utrecht University/Nikhef, Utrecht, Netherlands\\
$^{60}$ Institute of Experimental Physics, Slovak Academy of Sciences, Ko\v{s}ice, Slovak Republic\\
$^{61}$ Institute of Physics, Homi Bhabha National Institute, Bhubaneswar, India\\
$^{62}$ Institute of Physics of the Czech Academy of Sciences, Prague, Czech Republic\\
$^{63}$ Institute of Space Science (ISS), Bucharest, Romania\\
$^{64}$ Institut f\"{u}r Kernphysik, Johann Wolfgang Goethe-Universit\"{a}t Frankfurt, Frankfurt, Germany\\
$^{65}$ Instituto de Ciencias Nucleares, Universidad Nacional Aut\'{o}noma de M\'{e}xico, Mexico City, Mexico\\
$^{66}$ Instituto de F\'{i}sica, Universidade Federal do Rio Grande do Sul (UFRGS), Porto Alegre, Brazil\\
$^{67}$ Instituto de F\'{\i}sica, Universidad Nacional Aut\'{o}noma de M\'{e}xico, Mexico City, Mexico\\
$^{68}$ iThemba LABS, National Research Foundation, Somerset West, South Africa\\
$^{69}$ Jeonbuk National University, Jeonju, Republic of Korea\\
$^{70}$ Johann-Wolfgang-Goethe Universit\"{a}t Frankfurt Institut f\"{u}r Informatik, Fachbereich Informatik und Mathematik, Frankfurt, Germany\\
$^{71}$ Korea Institute of Science and Technology Information, Daejeon, Republic of Korea\\
$^{72}$ KTO Karatay University, Konya, Turkey\\
$^{73}$ Laboratoire de Physique Subatomique et de Cosmologie, Universit\'{e} Grenoble-Alpes, CNRS-IN2P3, Grenoble, France\\
$^{74}$ Lawrence Berkeley National Laboratory, Berkeley, California, United States\\
$^{75}$ Lund University Department of Physics, Division of Particle Physics, Lund, Sweden\\
$^{76}$ Nagasaki Institute of Applied Science, Nagasaki, Japan\\
$^{77}$ Nara Women{'}s University (NWU), Nara, Japan\\
$^{78}$ National and Kapodistrian University of Athens, School of Science, Department of Physics , Athens, Greece\\
$^{79}$ National Centre for Nuclear Research, Warsaw, Poland\\
$^{80}$ National Institute of Science Education and Research, Homi Bhabha National Institute, Jatni, India\\
$^{81}$ National Nuclear Research Center, Baku, Azerbaijan\\
$^{82}$ National Research and Innovation Agency - BRIN, Jakarta, Indonesia\\
$^{83}$ Niels Bohr Institute, University of Copenhagen, Copenhagen, Denmark\\
$^{84}$ Nikhef, National institute for subatomic physics, Amsterdam, Netherlands\\
$^{85}$ Nuclear Physics Group, STFC Daresbury Laboratory, Daresbury, United Kingdom\\
$^{86}$ Nuclear Physics Institute of the Czech Academy of Sciences, Husinec-\v{R}e\v{z}, Czech Republic\\
$^{87}$ Oak Ridge National Laboratory, Oak Ridge, Tennessee, United States\\
$^{88}$ Ohio State University, Columbus, Ohio, United States\\
$^{89}$ Physics department, Faculty of science, University of Zagreb, Zagreb, Croatia\\
$^{90}$ Physics Department, Panjab University, Chandigarh, India\\
$^{91}$ Physics Department, University of Jammu, Jammu, India\\
$^{92}$ Physics Program and International Institute for Sustainability with Knotted Chiral Meta Matter (SKCM2), Hiroshima University, Hiroshima, Japan\\
$^{93}$ Physikalisches Institut, Eberhard-Karls-Universit\"{a}t T\"{u}bingen, T\"{u}bingen, Germany\\
$^{94}$ Physikalisches Institut, Ruprecht-Karls-Universit\"{a}t Heidelberg, Heidelberg, Germany\\
$^{95}$ Physik Department, Technische Universit\"{a}t M\"{u}nchen, Munich, Germany\\
$^{96}$ Politecnico di Bari and Sezione INFN, Bari, Italy\\
$^{97}$ Research Division and ExtreMe Matter Institute EMMI, GSI Helmholtzzentrum f\"ur Schwerionenforschung GmbH, Darmstadt, Germany\\
$^{98}$ Saga University, Saga, Japan\\
$^{99}$ Saha Institute of Nuclear Physics, Homi Bhabha National Institute, Kolkata, India\\
$^{100}$ School of Physics and Astronomy, University of Birmingham, Birmingham, United Kingdom\\
$^{101}$ Secci\'{o}n F\'{\i}sica, Departamento de Ciencias, Pontificia Universidad Cat\'{o}lica del Per\'{u}, Lima, Peru\\
$^{102}$ Stefan Meyer Institut f\"{u}r Subatomare Physik (SMI), Vienna, Austria\\
$^{103}$ SUBATECH, IMT Atlantique, Nantes Universit\'{e}, CNRS-IN2P3, Nantes, France\\
$^{104}$ Sungkyunkwan University, Suwon City, Republic of Korea\\
$^{105}$ Suranaree University of Technology, Nakhon Ratchasima, Thailand\\
$^{106}$ Technical University of Ko\v{s}ice, Ko\v{s}ice, Slovak Republic\\
$^{107}$ The Henryk Niewodniczanski Institute of Nuclear Physics, Polish Academy of Sciences, Cracow, Poland\\
$^{108}$ The University of Texas at Austin, Austin, Texas, United States\\
$^{109}$ Universidad Aut\'{o}noma de Sinaloa, Culiac\'{a}n, Mexico\\
$^{110}$ Universidade de S\~{a}o Paulo (USP), S\~{a}o Paulo, Brazil\\
$^{111}$ Universidade Estadual de Campinas (UNICAMP), Campinas, Brazil\\
$^{112}$ Universidade Federal do ABC, Santo Andre, Brazil\\
$^{113}$ Universitatea Nationala de Stiinta si Tehnologie Politehnica Bucuresti, Bucharest, Romania\\
$^{114}$ University of Cape Town, Cape Town, South Africa\\
$^{115}$ University of Derby, Derby, United Kingdom\\
$^{116}$ University of Houston, Houston, Texas, United States\\
$^{117}$ University of Jyv\"{a}skyl\"{a}, Jyv\"{a}skyl\"{a}, Finland\\
$^{118}$ University of Kansas, Lawrence, Kansas, United States\\
$^{119}$ University of Liverpool, Liverpool, United Kingdom\\
$^{120}$ University of Science and Technology of China, Hefei, China\\
$^{121}$ University of South-Eastern Norway, Kongsberg, Norway\\
$^{122}$ University of Tennessee, Knoxville, Tennessee, United States\\
$^{123}$ University of the Witwatersrand, Johannesburg, South Africa\\
$^{124}$ University of Tokyo, Tokyo, Japan\\
$^{125}$ University of Tsukuba, Tsukuba, Japan\\
$^{126}$ Universit\"{a}t M\"{u}nster, Institut f\"{u}r Kernphysik, M\"{u}nster, Germany\\
$^{127}$ Universit\'{e} Clermont Auvergne, CNRS/IN2P3, LPC, Clermont-Ferrand, France\\
$^{128}$ Universit\'{e} de Lyon, CNRS/IN2P3, Institut de Physique des 2 Infinis de Lyon, Lyon, France\\
$^{129}$ Universit\'{e} de Strasbourg, CNRS, IPHC UMR 7178, F-67000 Strasbourg, France, Strasbourg, France\\
$^{130}$ Universit\'{e} Paris-Saclay, Centre d'Etudes de Saclay (CEA), IRFU, D\'{e}partment de Physique Nucl\'{e}aire (DPhN), Saclay, France\\
$^{131}$ Universit\'{e}  Paris-Saclay, CNRS/IN2P3, IJCLab, Orsay, France\\
$^{132}$ Universit\`{a} degli Studi di Foggia, Foggia, Italy\\
$^{133}$ Universit\`{a} del Piemonte Orientale, Vercelli, Italy\\
$^{134}$ Universit\`{a} di Brescia, Brescia, Italy\\
$^{135}$ Variable Energy Cyclotron Centre, Homi Bhabha National Institute, Kolkata, India\\
$^{136}$ Warsaw University of Technology, Warsaw, Poland\\
$^{137}$ Wayne State University, Detroit, Michigan, United States\\
$^{138}$ Yale University, New Haven, Connecticut, United States\\
$^{139}$ Yonsei University, Seoul, Republic of Korea\\
$^{140}$ Affiliated with an institute covered by a cooperation agreement with CERN\\
$^{141}$ Affiliated with an international laboratory covered by a cooperation agreement with CERN.\\

\end{flushleft} 

\end{document}